\documentclass[12pt,preprint]{aastex}
\shorttitle{Binary White Dwarf Merger Systems}
\shortauthors{KILIC ET AL.}

\begin{document}

\title{The ELM Survey. II. Twelve Binary White Dwarf Merger Systems$^\dagger$}

\author{Mukremin Kilic\altaffilmark{1,6},
Warren R. Brown\altaffilmark{1},
Carlos Allende Prieto\altaffilmark{2,3},
M. A. Ag\"{u}eros\altaffilmark{4},
Craig Heinke\altaffilmark{5},
and S. J. Kenyon\altaffilmark{1}}

\altaffiltext{$\dagger$}{Based on observations obtained at the MMT Observatory, a joint facility
of the Smithsonian Institution and the University of Arizona.}
\altaffiltext{1}{Smithsonian Astrophysical Observatory, 60 Garden St., Cambridge, MA 02138, USA; mkilic@cfa.harvard.edu}
\altaffiltext{2}{Instituto de Astrof\'{\i}sica de Canarias, E-38205, La Laguna, Tenerife, Spain}
\altaffiltext{3}{Departamento de Astrof\'{\i}sica, Universidad de La Laguna, E-38205, La Laguna, Tenerife, Spain}
\altaffiltext{4}{Columbia University, Department of Astronomy, 550 West 120th Street, New York, NY 10027, USA}
\altaffiltext{5}{Department of Physics, University of Alberta, 11322 - 89 Avenue, Edmonton, AB, T6G 2G7, Canada}
\altaffiltext{6}{\it Spitzer Fellow}

\begin{abstract}

We describe new radial velocity and X-ray observations of extremely low-mass 
white dwarfs (ELM WDs, $\sim0.2 M_{\odot}$) in the Sloan Digital Sky Survey Data Release 4 and
the MMT Hypervelocity Star survey. We identify four new short period binaries, including two
merger systems.
These observations bring the total number of short period binary systems identified in our survey to 20.
No main-sequence or neutron star companions are visible in the available optical photometry, radio,
and X-ray data. Thus, the companions are most likely WDs. Twelve of these systems will merge within a Hubble
time due to gravitational wave radiation. We have now tripled the number of
known merging WD systems.
We discuss the characteristics of this merger sample and potential links to underluminous supernovae,
extreme helium stars, AM CVn systems, and other merger products. We provide
new observational tests of the WD mass-period distribution and cooling
models for ELM WDs. We also find evidence for a new formation channel for single low-mass WDs
through binary mergers of two lower mass objects.

\end{abstract}


\section{INTRODUCTION}

Before the recent discovery of the six short period binary WD systems containing $\sim 0.2 M_{\odot}$
ELM WDs \citep{kilic10b,kilic10c,mullally09,marsh10,kulkarni10,badenes09,steinfadt10,kawka10},
there were only six double WD systems known to have short enough orbital periods to merge within a Hubble time
\citep[see][and references therein]{nelemans05}. Even though the Supernovae Ia Progenitor Survey (SPY)
identified $\approx$100 new double degenerate systems from a sample of 1000 WDs observed using the Very Large Telescope,
\citet{napiwotzki04} identify only eight merger candidates (including both WD+WD and sdB+WD systems) from the SPY and the
literature.
This fraction (8 out of 1000) is relatively small compared to the fraction of merger systems (6 out of 8)
among the ELM WDs in the literature \citep{kilic07b,kilic09,kilic10b,kilic10c}.

Low-mass ($M<0.45 M_{\odot}$) He-core WDs are formed when a companion strips the outer envelope from a post main-sequence star
before the star reaches the tip of the red giant branch and ignites helium. They are usually found
in close binaries, mostly double degenerate systems \citep{marsh95}. However, about half of the low-mass
WDs in the field do not show any radial velocity variations, indicating that they are single \citep{maxted00,napiwotzki07}.
\citet{kilic07c} argue that these single low-mass WDs may come from old metal-rich stars that truncate their evolution prior
to the helium flash from severe mass loss. They estimate a binary fraction of 50\% for
$\sim 0.4$~M$_\odot$ WDs. However, they predict that the binary fraction rises to 100\% for $\sim0.2 M_{\odot}$ ELM WDs,
since such extreme mass loss rates are not expected even for the most metal-rich stars in the Galaxy.

\subsection{The ELM Survey}

To study the binary fraction of ELM WDs and to find future merger systems, we have undertaken an MMT survey of all previously
identified ELM WDs from the SDSS DR4 area \citep{eisenstein06,liebert04} and
the Hypervelocity Star survey of \citet{brown06}. The latter sample is described in a companion paper by \citet{brown10a}.
Part of the SDSS DR4 sample was discussed earlier in \citet{kilic10b}. Here, we present radial velocity observations
of 7 more ELM WD candidates from the SDSS sample.
The combined sample contains 23 WDs. We also report on {\it Chandra} X-ray
observations of three ELM WDs to search for X-ray evidence of milli-second pulsar (MSP) companions.

While almost all SDSS ELM WDs show radial velocity variations, none show evidence of main-sequence or neutron star
companions in the optical, radio \citep{agueros09a}, or X-ray \citep[this paper and][]{agueros09b}.
Hence, they are almost certainly binary WD systems.
Out of the 20 short period binary systems in our sample, 12 have merger times shorter than a Hubble time.
Our ELM survey has now tripled the number of known merging WD systems.
Based on our sample and previously known short period binary WD systems, we study
the mass-period distribution and future evolutionary prospects, including the connections to unusual stellar populations and
underluminous supernovae (SNe).

Our paper is organized as follows. In Section 2 we discuss our
radial velocity and X-ray observations. In Section 3 we present an analysis of the optical spectroscopy and X-ray data and
the nature of the companions. In Section 4 we discuss the
sample characteristics of ELM WDs, including the binary fraction, mass-period distribution,
merging systems, and the implications for underluminous SNe and other
unusual merger products. We conclude in Section 5. 

\section{OBSERVATIONS}

\subsection{Optical Spectroscopy}

\citet{eisenstein06} identified 13 low-mass WD candidates with $\log g < 7$. \citet{kilic10b} discuss
follow-up observations of four of these targets. We selected 6 more targets for follow-up MMT spectroscopy.
These targets are SDSS J002228.45+003115.5, J002207.65$-$101423.5, J123410.36$-$022802.8, J162542.10+363219.1,
J204949.78+000547.3, and J225242.25$-$005626.6. We also observed SDSS J234536.48$-$010204.8, another ELM WD candidate
identified by \citet{liebert04}.  

There are three more targets in the \citet{eisenstein06} ELM WD sample that lack follow-up spectroscopy. These targets
(SDSS J1056+6536, J1426+0100, and J1630+4233) have $\log g>6.9$  based on the model fits by \citet{eisenstein06}
and $\log g\approx7$ based on the fits by \citet{kilic07a}. Radial velocity observations of these objects will be useful for
the identification of more short period binary WD systems.

We used the 6.5m MMT equipped with the Blue Channel Spectrograph to obtain moderate
resolution spectroscopy of seven ELM WD candidates over 17 different nights between 2008 September
and 2010 July.
We operate the spectrograph with
the 832 line mm$^{-1}$ grating in second order, providing wavelength coverage 3600
\AA\ to 4500 \AA\ and a spectral resolution of 1.0 - 1.2 \AA, depending on the slit size used.
We obtain all observations at the parallactic angle, with a comparison
lamp exposure paired with every observation. We flux-calibrate using blue
spectrophotometric standards \citep{massey88}, and we measure radial velocities
using the cross-correlation package RVSAO \citep{kurtz98}.
Our radial velocity measurement procedures are described in \citet{kilic09,kilic10b}.

\subsection{X-ray observations}

\subsubsection{Motivation}

X-ray emission from radio MSPs is expected to be less highly beamed
than radio emission, as a major component of the X-ray emission
comes from the hot neutron star surface \citep{zavlin02,zavlin06}.
Blackbody emission from the surface of a possible pulsar companion to
the ELM WDs will be gravitationally bent, allowing observation of
$>$75\% of the neutron star surface in X-rays even if the radio pulsar beam
(produced well above the surface) misses our line of sight
\citep{beloborodov02}.
All 15 radio MSPs with precise positions in unconfused regions of
the globular cluster 47 Tuc have been clearly detected in X-rays
\citep{heinke05,bogdanov06}, with no observed correlation between
their radio and X-ray luminosities (expected, as they are produced via
different mechanisms).  There is also no evidence
for differences in the X-ray properties of globular
cluster vs. Galactic MSPs \citep{bogdanov06}.

This result allows us to use the 47 Tuc
MSP sample (with accurate $L_X$ values due to its well-known distance)
to predict that other MSPs
should have X-ray luminosities above $L_X$(0.5-6 keV)$=2\times10^{30}$
erg/s, the minimum $L_X$ of MSPs in 47 Tuc. 
Thus, deep X-ray upper limits can rule out the
presence of an MSP, as in the ELM WD SDSS J0917+4638 \citep{agueros09b}.
The MSPs in 47 Tuc show X-ray spectra dominated by thermal
blackbody-like emission from the neutron star polar cap surfaces, with
temperature $1-3\times10^6$ K \citep{bogdanov06}.  Some MSPs
also show additional (sometimes much higher $L_X$) components due to
magnetospheric emission or pulsar wind shocks, but there is no
evidence for a MSP lacking the thermal component.

We obtained {\it Chandra} observations of three ELM WDs (SDSS
J105353.89+520031.0, J123410.36$-$022802.8, and J092345.59+302805.0)
to search for MSP X-ray
emission.  None of these WDs had been previously observed in
the X-rays since the {\it ROSAT} All-Sky Survey \citep{voges99}, where the
short exposure time was insufficient to place useful limits.

\subsubsection{Data Analysis}

We observed the three systems with {\it Chandra}'s ACIS-S instrument
in Very Faint mode, for the exposure times noted in Table 1.
We used CIAO 4.2 (with CALDB
4.2.1)\footnote{http://cxc.harvard.edu/ciao/} to reprocess the
observations with current calibrations and reduce the backgrounds
using Very Faint mode cleaning. We constructed images in the 0.3-6
keV band, and found no X-ray photons within the 1'' error circles
around each source.  We use Poisson statistics \citep{gehrels86} to place
99\% confidence upper limits on the count rate.  We compute distances
to the ELM WDs using the models of \citet{panei07} and the SDSS $g$
magnitudes, and $N_H$ values using the Colden 
tool\footnote{http://asc.harvard.edu/toolkit/colden.jsp}
\citep{dickey90}. We use
PIMMS\footnote{http://asc.harvard.edu/toolkit/pimms.jsp} and the X-ray
spectrum of the faintest MSP in 47 Tuc (47 Tuc-T, 134 eV blackbody) to
produce 0.5-6 keV $L_X$ upper limits, which we list in Table 1.

The 99\% confidence $L_X$ upper limits are all at least a factor of 10
fainter than the faintest MSP identified in 47 Tuc, and
21 times fainter than the median $L_X$ of these MSPs
\citep{bogdanov06}.  Thus the lack of X-ray emission from these WDs
is strong evidence that the unseen companions are not MSPs.

\section{RESULTS}

Four of our seven targets exhibit significant radial velocity variations, with peak-to-peak velocity amplitudes
between 124 and 232 km s$^{-1}$. 
We weight each velocity measurement by its associated error and solve for the best-fit
orbit using the code of \citet{kenyon86}. 
Figures 1-6 show the observed radial velocities and the best fit orbits for our targets.
The heliocentric radial velocities are best fit
with circular orbits with periods 1.9-11.8 hr. 
We present the best-fit orbital period ($P$), semi-amplitude
($K$) of the radial velocity variations, systemic velocity (which includes a small gravitational redshift term),
the time of spectroscopic conjunction, and the mass function in Table 2. J2049+0005 is excluded from this list due
to its low surface gravity, which implies that it is not a WD (see below).

Two of the stars in our sample, J2252$-$0056 and J2345$-$0102, do not show significant velocity variations. The best-fit
orbital solutions have semi-amplitudes 25-43 km s$^{-1}$. These two objects
may be low inclination systems or just single stars.

We perform model fits to each individual spectrum and also to the average composite spectra using synthetic DA WD spectra kindly
provided by D. Koester. This model grid covers temperatures up to 30,000 K. We use an NLTE grid computed by I. Hubeny
\citep[see][]{allende09} to fit the spectrum of J2345$-$0102, which requires a solution slightly above 30,000 K.
We use the individual spectra to obtain a robust estimate of the errors in our analysis.
Figure 7 shows the composite spectra and our fits using the entire wavelength range. The best-fit $T_{\rm eff}$ and
$\log g$ values are given in Table 3.
We obtain best-fit solutions of $17,890-33,130$ K and $\log g= 6.12-7.38$ for six of our targets,
confirming that they are low-mass WDs.

Figure 8 shows the effective temperatures and surface gravities for our targets (red circles) plus the previously
identified ELM WDs in the literature \citep{kilic07b,kilic09,vennes09,kilic10b,brown10a}. Filled triangles show the
WD companions to milli-second pulsars PSR J1012+5307 and J1911-5958A
\citep[][see \S 4.4]{vankerkwijk96,bassa06}. 
Solid lines show the constant mass tracks for low mass WDs from \citet{panei07} models \citep[updated by][]{kilic10b}.
This figure shows that our six WD targets have masses ranging from 0.20 to 0.42 $M_{\odot}$.

\subsection{Notes on Individual Objects}

\subsubsection{J0022+0031}

The best-fit orbital period for J0022+0031 is 11.8 hr. However, the limited number of follow-up observations allow
for an alias at 7.9 hr period.
J0022+0031 is a binary system containing a 120 Myr old 0.38 $M_{\odot}$ WD and a $M\geq0.21 M_{\odot}$ companion star.
Based on the updated \citet{panei07} models, it has an absolute magnitude of $M_{\rm g}=9.8$ mag and a distance
of 790 pc.

For all six WDs in our sample, we combine the spectra near maximum blue-shifted and red-shifted velocities into
two composite spectra. If there is a contribution from a companion object, it may be visible as an asymmetry in the line profiles.
We do not see any obvious asymmetries in the line profiles and our optical spectroscopy does not reveal any spectral
features from companion objects. A $M\geq0.21~M_\odot$ main-sequence star companion to J0022+0031
would have $M_I\leq10.0$ mag \citep{kroupa97}, brighter than the low-mass WD ($M_I\approx 10.4$ mag) and detectable in the $I$-band.
Such a main-sequence star companion is ruled out based on the SDSS photometry.

Using the mean inclination angle for a random stellar sample, $i=60^{\circ}$, the companion has a mass of 0.26 $M_{\odot}$. 
There is a 2.5\% probability that the companion is a 1.4-3 $M_{\odot}$ neutron star. Given the small probability of a neutron star companion
and the unsuccesful searches for neutron stars around other ELM WDs \citep[Section 2.2, see also][]{agueros09a,agueros09b}, the companion is most
likely another WD. Short period binaries may lose angular momentum through gravitational wave radiation and merge within a Hubble time.
The merger time for the J0022+0031 system is longer than a Hubble time.

\subsubsection{J0022$-$1014}

J0022$-$1014 is a 1.9 hr period binary system with a 0.33 $M_{\odot}$ WD primary and a $M\geq0.19 M_{\odot}$ secondary. The primary
WD is 70 Myr old, it has $M_{\rm g}$ = 9.2 mag, and is located at a distance of 1.3 kpc. 
A $M\geq0.19 M_{\odot}$ main-sequence star companion would have $M_I\leq10.3$ mag, 30\% fainter than the ELM WD. Such a companion
is ruled out based on the SDSS photometry. The companion to J0022$-$1014 is most likely another WD, and specifically a low-mass
WD. There is a 83\% probability that the companion is less massive than 0.45 $M_{\odot}$, a He-core low-mass WD. This system will
merge in less than 730 Myr. 

J0022$-$1014 has a proper motion of $\mu_{\alpha} cos \delta = -7.8$ and $\mu_{\delta}=-13.2$ mas yr$^{-1}$
\citep{munn04}. Based on the mass and radius estimates, the gravitational redshift of the WD
is 8.1 km s$^{-1}$. Therefore, the true systemic velocity is $-46.6$ km s$^{-1}$.
The velocity components with respect to the local standard of rest as defined by \citet{hogg05} are
$U = 95 \pm 23, V = -53 \pm 21$, and $W = 33 \pm 8$ km s$^{-1}$. J0022$-$1014 is
a disk star.

\subsubsection{J1234$-$0228}

J1234$-$0228 was originally identified as a very low mass WD by \citet{liebert04} and is included in the DR4 WD
catalog of \citet{eisenstein06}. Based on our MMT observations over 5 different nights separated by about a year,
J1234$-$0228 is a binary system with an orbital period of 2.2 hr. The visible component of the binary is a
70 Myr old WD with $M_{\rm g}$ = 8.4 mag at a distance of 780 pc. For an inclination angle of 60$^{\circ}$, the mass
function requires a 0.11 $M_{\odot}$ companion. A main-sequence companion of this mass would contribute 7\% excess
flux in the $I-$band.
No excess flux is observed in the SDSS photometry and spectrum of this object, therefore the companion is most likely another
WD \citep[see also][]{liebert04}. 

There is a 94\% chance that the companion is a low-mass ($<0.45 M_{\odot}$) WD, and more interestingly there is a 85\%
probability that the combined mass of the two stars in this system is lower than 0.45 $M_{\odot}$. The merger time for
the J1234$-$0228 system is shorter than 2.7 Gyr. Therefore, within the next few Gyr, this system will merge and likely form
a single low-mass WD.

J1234$-$0228 has a proper motion of $\mu_{\alpha} cos \delta = -14.5$ and $\mu_{\delta}=-12.3$ mas yr$^{-1}$
\citep{munn04}. Based on the mass and radius estimates, the gravitational redshift of the WD
is 3.7 km s$^{-1}$. Therefore, the true systemic velocity is +46.6 km s$^{-1}$.
The velocity components with respect to the local standard of rest as defined by \citet{hogg05} are
$U = -5 \pm 14, V = -78 \pm 12$, and $W = 21 \pm 7$ km s$^{-1}$. J1234$-$0228 is also a disk star.

\subsubsection{J1625+3632}

J1625+3632 has a best-fit orbital period of 5.6 hr. The MMT spectrum reveals a weak \ion{He}{1} line at 4471 \AA.
Our best-fit DA WD model has $T_{\rm eff}=$ 23570 K and $\log g=$ 6.12, similar to the sdB star HD 188112
\citep{heber03}. Based on these model fits, the primary is a $\approx$160 Myr old 0.20 $M_{\odot}$ WD at
a distance of 2.8 kpc. J1625+3632 does not display a significant proper motion,
but given its distance and systemic velocity, it is likely a member of the thick disk or halo.

The relatively small semi-amplitude ($K= 58.4$ km s$^{-1}$) of the radial velocity variations imply that
the lower limit on the mass of the companion is 0.07 $M_{\odot}$. A main-sequence companion of this mass would contribute less than
1\% in the optical and would not be detected in our observations or the SDSS photometry. Given the accuracy of the SDSS photometry,
any main-sequence companion more massive than about 0.17 $M_{\odot}$ would have been detected. Hence, only 0.07-0.17 $M_{\odot}$
main-sequence companions are allowed based on the mass function and optical photometry. Near-infrared observations will be useful
to rule out such companions.

J1625+3632 may be a low inclination system. However, the probability of a neutron star companion is
only 0.6\%. Given the narrow range of possible main-sequence companions and the unlikelihood of a neutron star companion,
the companion is likely another WD. There is a 96\% chance that the companion is less massive than 0.45 $M_{\odot}$, i.e., a low-mass
He-core WD.

\subsubsection{J2049+0005}

Figure 9 shows the best-fit WD model to our MMT spectrum of J2049+0005. This star was identified as a
$T_{\rm eff}$ = 8660 K and $\log g$ = 5.48 WD by \citet{eisenstein06}. Further analysis of the same
SDSS spectrum by \citet{kilic07a} suggested that J2049+0005 has $\log g<5$, and therefore is not a WD.
Similarly, our model fits to the higher resolution and higher signal-to-noise ratio MMT spectrum find
a best-fit solution that is exactly our lowest gravity model ($\log g$ = 5). This model poorly matches
the high order Balmer lines, indicating that the surface gravity is lower than $\log g$ = 5 and that
J2049+0005 is most likely an A star.

\subsubsection{J2252$-$0056}

Based on 30 different spectra, the best-fit orbital period is 10.3 hr with $K= 25.1 \pm 1.5$ km s$^{-1}$. However,
the radial velocity variations are not significant. The best-fit orbit has a $\chi^2$ slightly better than a
constant-velocity fit. If J2252$-$0056 is a binary, it is likely a low inclination or a pole-on system.
Given the 25 km s$^{-1}$ upper limit to the velocity semi-amplitude,
J2252$-$0056 could be a binary with either month-long (assuming an 0.5 $M_{\odot}$ companion and $i=60^\circ$) orbital period 
or relatively pole-on inclination ($M \sin{(i)} = 0.03-0.05 M_{\odot}$ for 2-12 hr periods).
There is no evidence of a companion in our MMT spectrum or the SDSS observations. 

\subsubsection{J2345$-$0102}

J2345$-$0102 is another object for which we detect no significant velocity variations. The best-fit orbital
period is 7.2 hr with $K= 43.0 \pm 0.7$ km s$^{-1}$, but the best-fit orbit is a slightly better fit to the data
than a constant-velocity fit. Again, if J2345$-$0102 is a binary, it may be a low inclination system.
The observed 43 km s$^{-1}$ upper limit to the velocity semi-amplitude corresponds to
a binary with either month-long orbital period or relatively pole-on inclination
($M \sin{(i)} = 0.05-0.10 M_{\odot}$ for 2-12 hr periods).

The MMT spectrum of this object is best-explained with a 23 Myr old 0.42 $M_{\odot}$ WD at a distance of 2.0 kpc.
Based on 5 epochs from the USNO-B+SDSS, \citet{munn04} measure a proper motion of 
$\mu_{\alpha} cos \delta = -26.0$ and $\mu_{\delta}=-36.6$ mas yr$^{-1}$. Given its relatively large systemic 
velocity, proper motion, and distance, J2345$-$0102 is clearly a halo star.
About half of the known 0.4 $M_{\odot}$ WDs seem to be single \citep[see the discussion in][]{kilic07c}. Hence,
the lack of significant radial velocity variations is not very surprising for this system.

\section{DISCUSSION}

\subsection{The Binary Fraction of ELM WDs}

\citet{kilic07b,kilic10b}, \citet{brown10a}, and this paper discuss radial velocity observations of 11 ELM WD candidates identified
in the SDSS DR4 \citep{eisenstein06} and 12 ELM WDs identified in the Hypervelocity Star Survey of \citet{brown06}.
One of the targets is common to both samples. In addition, one of the DR4 targets (J2049+0005) is most likely an A-star.
\citet{kilic09,kilic10c}, \citet{kawka10}, and \citet{steinfadt10} describe radial velocity observations of two more ELM WDs
identified in the literature. Thus, there are 23 (candidate) ELM WD systems with comprehensive radial velocity measurements.
All but three of these systems (J1448+1342, J2252$-$0056, and J2345$-$0102) are in binary sytems.

Limiting this sample to 19 WDs with $M\leq0.25 M_{\odot}$,
18 are in short period binary systems. Thus, close binary evolution is required for the formation of ELM WDs.
The only system that does not display significant radial velocity variations, J1448+1342, is at the upper mass limit of
this selection; all 18 WDs with $M<0.25 M_{\odot}$ are indeed binaries.
The average velocity semi-amplitude of these 18 binaries is 236 km s$^{-1}$, whereas the upper limit for
the velocity semi-amplitude of J1448+1342 is 35 km s$^{-1}$. An average system viewed at $i\leq8.5^\circ$ 
would be consistent with the observations of J1448+1342.
For a randomly distributed sample of orbital inclinations, there is a 1.1\% chance that $i\leq8.5^\circ$.
Thus, there is a 21\% likelihood of finding one of the 19 systems with $i\leq8.5^\circ$.
It is possible that J1448+1342 is a pole-on binary system \citep[see also][]{brown10a}.
Hence, the binary fraction of $M\leq0.25 M_{\odot}$ WDs is at least 95\% and it may be as high as 100\%.

This result is in contrast to the binary fraction of slightly more massive WDs. \citet{maxted00}, \citet{napiwotzki04},
and \citet{brownjm10} demonstrate that about 30-50\% of $M\approx0.4 M_{\odot}$ WDs are single \citep[see also][]{kilic10a}.
Single low-mass WDs may form through enhanced mass loss from a metal-rich progenitor star \citep{kilic07c}, but we do not
expect even the most metal-rich stars to loose enough mass to create $\sim 0.2 M_{\odot}$ WDs. Here we confirm this expectation
that the lowest mass WDs all form in close binary systems. 

\subsection{A Dozen Merger Systems}

Double degenerate (DD) merger systems are one of the two main accepted formation channels for Type Ia supernovae
\citep[][and references therein]{iben84,branch95,kotak08,distefano10}.
However, only a handful of DD systems are known to have combined mass above the Chandrasekhar limit and
short enough orbital periods to merge through gravitational wave radiation within a Hubble time.
\citet{napiwotzki01} started the heroic SPY survey for short period DD systems. They used the Very Large Telescope to
observe 1014 stars brighter than 16.5 mag, including 75\% of the known WDs accessible from the southern hemisphere.
\citet{napiwotzki04} report that the SPY survey discovered about 100 new DD systems, one of which is a potential
SNe Ia progenitor \citep[WD 2020$-$425,][]{napiwotzki07}. \citet{geier10} describe the discovery of another massive
DD merger system and also a summary of the current discoveries from the SPY survey (their Figure 4).

Figure 10 shows the total masses and periods for the short period DD systems (including sdB+WD) discovered by the SPY survey
(kindly made available to us by S. Geier) and our ELM WD sample. To explain the observed SNe Ia rate, only
one out of 1000 WDs is expected to
be a SNe Ia progenitor \citep{nelemans01}. Hence, the relatively few number of potential SNe Ia progenitors
in the SPY survey is not surprising. What is striking is that the number of merger systems found in the SPY survey
is significantly smaller than our ELM WD survey. We discovered 12 binary WD merger systems
($\tau_{\rm merge}<\tau_{\rm hubble}$) in a sample
of 23 stars, whereas the SPY survey discovered a handful of such systems in a sample of 1000 stars.
The reason for this is the binary common envelope evolution required to form ELM WDs.

Figure 11 shows the total masses and periods for previously identified WD+WD binary systems in the literature
\citep{nelemans05,napiwotzki07}. There were only six double WD merger systems known prior to our work. With the additional
12 merger systems presented in Table 4, we now have tripled the number of WD merger systems known. 

\subsection{The Period Distribution of Binary WDs}

\citet{nelemans01} studied the period and mass distribution of double WD systems using population synthesis models.
The details of these models including selection effects due to orbital evolution of the systems (disappearence
of the shortest period systems due to mergers), target selection biases in the current surveys, and different cooling
models are discussed in their paper. \citet{nelemans01} find a correlation between the masses and periods; more massive
WD primaries have longer periods. This is not surprising; shorter period systems would start interacting earlier
in their evolution compared to longer period systems and experience enhanced mass-loss during the red giant phase, hence
end up as lower mass WDs. They also find that the mass ratio distribution peaks at $q=1$, favoring equal mass
binary components. This distribution is consistent with the observed mass ratio distribution of binary WDs known prior to our work.

Figure 12 shows the masses and orbital periods for our sample of ELM WDs and the previously known binary WD systems
in the literature compared to the population synthesis models of \citet[][kindly made available
to us by G. Nelemans]{nelemans01,nelemans04}. Depending on which cooling model is used, the model predictions change somewhat
\citep[see][for a discussion of the cooling models used here]{nelemans04}. However,
the main trend in the mass versus period distribution is evident; lower mass primaries are in
shorter period systems. This result is consistent with the observed distribution of masses and periods for previously
known binary WD systems and the ELM WD systems. All but one of the $M<0.3 M_{\odot}$ systems have
periods shorter than a day\footnote{The single velocity non-variable system, J1448+1342, is probably
a pole-on or a relatively long-period binary system.},
whereas the majority of the more massive WDs have periods longer than a day.

The overall number distribution of the population synthesis models and the observations cannot be directly compared
without taking into account the obvious biases in the target selection.
The sample of previously known binary WDs are selected from a
magnitude limited sample down to 16-17 mag, whereas our sample mainly comes from fainter targets selected from the SDSS,
which has its own spectroscopic target selection bias. Nevertheless, an overabundance of 0.17 $M_{\odot}$ WDs is evident
in Figure 12. A comparison between the population synthesis models using \citet[][top panel]{hansen99} and the modified
\citet[][bottom panel]{driebe98} cooling models shows that the expected number of the lowest mass WDs changes depending on
the cooling model used.
The latter models predict a significantly larger number of 0.17 $M_{\odot}$ WDs. Based
on the \citet{panei07} models, a 0.25 $M_{\odot}$ WD cools 4$\times$ faster than a 0.17 $M_{\odot}$ WD down to an
absolute magnitude of $M_g=10$ mag. Hence, the observed mass distribution of ELM WDs in a magnitude-limited survey
is skewed toward lower mass objects \citep{brown10b}. Population synthesis models and a large, unbiased, magnitude-limited
survey of low-mass and ELM WDs like that of \citet{brown10a} will
be useful to constrain the ELM WD cooling models.

Another important prediction of the population
synthesis models is that there should not be many systems with periods less than an
hour\footnote{These systems either merge or evolve into AM CVn systems relatively quickly.}.
We have three ELM WD systems with periods close to 1 hr. The shortest period system, J1053+5200
has a period of 61 min \citep{kilic10b,mullally09}.
The shortest period hydrogen-rich cataclysmic variable (CV) V485 Cen has an orbital period
of 59 min \citep{augusteijn96}.
Both short period binary WDs and CVs are expected to contribute to the
AM CVn population, which contains 
interacting WD binary systems with orbital periods ranging from a few minutes
to about an hour. The two longest period AM CVn systems have periods of 56 min and 66 min, respectively \citep{roelofs07}.
It seems that hydrogen-deficient AM CVn interacting binary systems have periods less than about an hour and the hydrogen-rich 
detached WD systems have orbital periods about an hour or longer.

\subsection{The Minimum Mass for Low-mass WDs}

Before the SDSS, ELM WDs were traditionally found as companions to MSPs.
The masses for the MSP companions are usually measured from the pulsar timing measurements since the WDs are usually too faint
for follow-up optical spectroscopy. \citet{vankerkwijk96} and \citet{bassa06} obtained Keck and VLT spectroscopy of companions to
two MSP systems. \citet{vankerkwijk96} derive $T_{\rm eff}=$ 8550 $\pm$ 25 K and $\log$ g = 6.75 $\pm$ 0.07 for the companion
to PSR J1012+5307, whereas \citet{bassa06} obtain $T_{\rm eff}=$ 10,090 $\pm$ 150 K and $\log$ g = 6.44 $\pm$ 0.20 for the
companion to PSR J1911-5958A. These two WDs are included in Figure 8.

\citet{kilic07a,kilic07b} discovered the lowest surface gravity WD known at the time, SDSS J0917+4638, with $\log g=5.55$.
\citet{brown10a} discovered two WDs with even lower surface gravities, SDSS J1233+1602 and J2119$-$0018. These WDs are
included in our ELM WD sample and in Figure 8. Based on the \citet{serenelli01,serenelli02} and \citet{panei07} models,
all of these WDs, including the two MSP companions, have $M\geq0.17 M_{\odot}$. There are about a dozen WDs in Figure 8
with $M\approx0.17 M_{\odot}$, but none below that mass limit. However, this result is most likely due to observational
biases present in the SDSS and the Hypervelocity Star Survey. The former is, of course, an incomplete spectroscopic survey and
the latter has a color selection that excludes $T_{\rm eff}\leq$ 10000 K objects with $\log g=5-6$ \citep[see Figure 1 in][]{brown10a}. Based
on the \citet{panei07} models, a 0.16 $M_{\odot}$ WD takes 15 Gyr to cool down from 9200 K to about 7000 K with an increase
in surface gravity from $\log g =$ 5.7 to 6.4 and a decrease in luminosity from $M_g=$ 8.1 to 11.1 mag.
Such WDs are not included in the Hypervelocity Star Survey color selection.

Based on radio timing measurements, 45 out of 99 MSPs with periods less than 20 ms
have companions less massive than $0.17 M_{\odot}$ (B. K{\i}z{\i}ltan 2010, private communication).
The spectral types of the companions are unknown. However, a large
fraction of these objects are expected to be low-mass WDs; nature produces $M<0.17 M_{\odot}$ WDs in MSP systems.
Lower mass ELM WDs should be produced in binary WD systems as well. Population synthesis calculations
for double WDs do not predict many WDs below 0.17 $M_{\odot}$, but such WDs may eventually be found
in double WD systems in the SDSS or other spectroscopic surveys.

\subsection{Merger Products: (underluminous) SNe Ia, AM CVn, and RCrB stars}

The end result of the future evolution of the 12 ELM WD merger systems in our sample depends
on the mass ratios of the two components. The systems with extreme mass ratios ($q\leq0.2$) should
have stable mass transfer at initial Roche lobe filling.
Stable mass transfer leads to the formation of AM CVn systems that transfer angular momentum
back to the orbit, which increases the orbital period. If the higher mass WDs in such systems have
masses close to the Chandrasekhar limit, accretion may result in a Type Ia SNe explosion. However,
the probability of such an event is small ($\leq$ 7\%) for our systems. This probability
would be higher if sub-Chandrasekhar mass WD explosions also contribute to the observed Type Ia SNe population
\citep[][and references therein]{sim10,kromer10}. 
However, our ELM survey is biased against SNe Ia progenitors due to the selection of very low-mass primaries.

Figure 13 shows the mass ratios and periods for our ELM WD sample plus the previously known
binary WD systems. Single-lined systems are shown as filled triangles, whereas the double-lined systems
and the eclipsing WD system NLTT 11748 are shown as filled circles. Open circles mark the objects
with merger times shorter than a Hubble time. Out of the 18 merger systems, six have mass ratios
smaller than 0.25 (assuming an average inclination of 60$^{\circ}$), and are likely to form
AM CVn systems. These six systems also have the highest total mass ($M_1 + M_2$) in our sample.
J1233+1602 is the most extreme case with $M_1 + M_2 = 1.37 M_{\odot}$ for $i=60^{\circ}$.
It is striking that the number of extreme mass ratio systems in our sample (six) is the same as the number
of AM CVn systems found in the SDSS Data Relase 5 area \citep{roelofs07}.
These systems are potential SNe .Ia progenitors \citep{bildsten07}.

Based on a magnitude-limited sample of six ELM WD mergers, \citet{brown10b} find that half of their sample
has mass ratios of $\approx$0.15 and they estimate a contribution factor of $\ga$3\% to the AM CVn formation rate
from the observed population of ELM WDs. The fraction of extreme mass ratio systems in our survey is also 50\%
(six out of 12 ELM WDs), supporting the conclusions reached by \citet{brown10b}.

The other six ELM WD systems in our sample have mass ratios larger than 0.4. 
\citet{kilic10a} discuss various evolutionary scenarios for these objects.
These systems will
merge within a Hubble time and form a variety of exotic objects including underluminous SNe, extreme helium
stars (RCrB), or single helium-enriched subdwarf O stars. The merger rate of ELM WDs
is roughly the same as the observed rate of underluminous SNe \citep{brown10b}.

SDSS J1234$-$0228 is a unique ELM WD merger system. There is a 85\%
chance that the combined mass of the two stars in this system is lower than 0.45 $M_{\odot}$. 
Therefore, within the next few billion years, this system will merge and likely form a
a single low-mass WD. There are several scenarios to explain the presence
of single low-mass ($\sim0.4M_{\odot}$) WDs, including enhanced mass-loss from metal-rich progenitors \citep{kilic07c} and mass loss
through common envelope
evolution with a low-mass brown dwarf or planetary mass companion \citep{nelemans98}.
Single low-mass WDs can also form through mergers of two lower-mass WDs \citep{iben97,saio00}, though
no such progenitors have been found until now.
Here we have uncovered a potential progenitor system for a future single low-mass WD, J1234$-$0228.

About 10\% of DA WDs
are low-mass \citep{liebert05} and half of these low-mass WDs are single \citep{napiwotzki04},
corresponding to a space density of 2.5 $\times 10^5$ kpc$^{-3}$ \citep{sion09}.
The SDSS DR4 WD catalog covers 4783 square degrees with $g=15-20$ mag \citep{eisenstein06}.
We estimate the local space density using the modified $1/V_{\rm max}$ method and the disk
model of \citet{roelofs07}.
Based on one system, J1234$-$0228, the space density of merging WD systems that can form single low-mass WDs is
1.1 kpc$^{-3}$. This number is likely uncertain by a factor of 10 due to incompleteness
of the SDSS spectroscopic target selection.
Nevertheless, it is several orders of magnitude smaller than the estimated space density of
single low-mass WDs. Hence, this formation channel does not significantly
contribute to the single low-mass WD population in the solar neighborhood.

\section{CONCLUSIONS}

Our MMT ELM survey has now discovered 20 short period binary WD systems, including 12 merger systems. Only six binary WD merger
systems were known prior to our work. The ELM survey has now tripled the number of such systems.
These merger systems provide potential formation channels for a variety of interesting objects including underluminous SNe,
AM CVn, extreme helium stars, single subdwarf stars, and now single low-mass WDs.

All but one of the 19 stars with $M\leq0.25 M_{\odot}$ are in binaries, strongly arguing for a binary formation channel
for ELM WDs.
The combined sample of more than 40 short period WD systems found in our ELM survey and the literature demonstrate
that the shortest period detached WD systems have $\approx$1 hr orbital periods and they also have the lowest mass primaries.
In addition, the lowest mass WDs in double WD systems have $M=0.17 M_{\odot}$. Lower-mass objects
may be produced in such systems but are currently undetected in the SDSS and other spectroscopic surveys.

The sample of ELM WD merger systems will be significantly enlarged by our continuing ELM
survey based on SDSS photometry. Such a magnitude-limited survey
will not suffer from the biases present in our current sample, and will provide invaluable information
on the merger rates of ELM WD systems and their contribution to underluminous SNe and AM CVn systems.
Furthermore, a large sample of ELM WDs and population synthesis models will improve our understanding of the cooling
models for ELM WDs.

\acknowledgements
We thank G. Nelemans for providing his binary WD population synthesis model results, S. Geier for
the latest results from the SPY survey, and D. Koester for WD model spectra.
Support for this work was provided by NASA through the {\em Spitzer Space Telescope} Fellowship Program,
under an award from Caltech.

{\it Facilities:} \facility{MMT (Blue Channel Spectrograph), CXO (ACIS-S)}

\clearpage

\begin{deluxetable}{lcccccc}
\tabletypesize{\footnotesize}
\tablewidth{0pt}
\tablecaption{X-ray Observations of ELM WDs}
\tablehead{
{Name} & {ObsID} & {Expos.} & {Dist.} & {$N_H$} &  {Countrate} & {$L_X$}\\
(SDSS) &         &  (ks)    & (kpc)   & (cm$^{-2}$)& (cts s$^{-1}$) & (ergs s$^{-1}$)
}
\startdata
J105353.89+520031.0 & 9963 & 20.50 & 1.08  & $1.0\times10^{20}$ &
$<2.2\times10^{-4}$ & $<9.5\times10^{28}$ \\
J123410.36$-$022802.8 & 9964 & 5.50   & 0.78  & $2.2\times10^{20}$ &
$<8.4\times10^{-4}$ &  $<2.0\times10^{29}$ \\
J092345.59+302805.0 & 9965 & 1.03   & 0.27 & $2.0\times10^{20}$ &
$<4.5\times10^{-3}$ &  $<1.3\times10^{29}$ \\
\enddata
\tablecomments{99\% confidence X-ray count rate upper limits for three
  ELM WDs from {\it Chandra} X-ray observations.
  Countrate limit is in 0.3-6 keV band and $L_X$ limit is in 0.5-6 keV band.
}
\end{deluxetable}

\begin{deluxetable}{ccrrcc}
\tabletypesize{\footnotesize}
\tablecolumns{6}
\tablewidth{0pt}
\tablecaption{Orbital Parameters}
\tablehead{
\colhead{Object}&
\colhead{$P$}&
\colhead{$K$}&
\colhead{$V_{\rm systemic}$}&
\colhead{Spec. Conjunction}&
\colhead{Mass Function}\\
  & (days) & (km s$^{-1}$) & (km s$^{-1}$) & (HJD $-$ 2454000) & ($M_{\odot}$)
}
\startdata
J0022$+$0031 & 0.4914 $\pm$ 0.0254 & 80.8  $\pm$  1.3  & $-$20.3 $\pm$  0.8 & 732.71818 & 0.02681 $\pm$ 0.00190 \\
J0022$-$1014 & 0.0799 $\pm$ 0.0030 & 145.6 $\pm$  5.6  & $-$38.5 $\pm$  3.7 & 732.76530 & 0.02553 $\pm$ 0.00310 \\
J1234$-$0228 & 0.0914 $\pm$ 0.0040 & 94.0  $\pm$  2.3  & $+$50.3 $\pm$  1.8 & 917.75366 & 0.00787 $\pm$ 0.00067 \\
J1625$+$3632 & 0.2324 $\pm$ 0.0396 & 58.4  $\pm$  2.7  & $-$95.0 $\pm$  2.1 & 922.74700 & 0.00480 $\pm$ 0.00105 \\
J2252$-$0056 & \nodata               & \nodata           & $-$23.2 $\pm$  0.8 & \nodata   & \nodata               \\
J2345$-$0102 & \nodata               & \nodata           & $-$161.2 $\pm$ 0.4 & \nodata   & \nodata               
\enddata
\tablecomments{J0022+0031 has a period alias at 7.9 hr.}
\end{deluxetable}

\begin{deluxetable}{ccccccccc}
\tabletypesize{\footnotesize}
\tablecolumns{9}
\tablewidth{0pt}
\tablecaption{Physical Parameters}
\tablehead{
\colhead{Object}&
\colhead{$T_{\rm eff}$}&
\colhead{$\log g$}&
\colhead{Mass}&
\colhead{$M_2$}& 
\colhead{$M_2(i=60^\circ)$}&
\colhead{NS}&
\colhead{SN Ia}&
\colhead{$\tau_{\rm merge}$}\\
  & (K) &  & ($M_\odot$) & ($M_\odot$) & ($M_\odot$) & Prob. & Prob. & (Gyr)
}
\startdata 
J0022$+$0031 & 17890 $\pm$ 110 & 7.38 $\pm$ 0.02 & 0.38 & $\geq 0.21$ & 0.26 & 2.5\% & 1.9\% & $\leq 75.7$ \\
J0022$-$1014 & 18980 $\pm$ 380 & 7.15 $\pm$ 0.04 & 0.33 & $\geq 0.19$ & 0.23 & 2.3\% & 1.4\% & $\leq 0.73$ \\
J1234$-$0228 & 18000 $\pm$ 170 & 6.64 $\pm$ 0.03 & 0.23 & $\geq 0.09$ & 0.11 & 0.9\% & 0.3\% & $\leq 2.69$ \\
J1625$+$3632 & 23570 $\pm$ 440 & 6.12 $\pm$ 0.03 & 0.20 & $\geq 0.07$ & 0.08 & 0.6\% & 0.2\% & $\leq 45.2$ \\
J2252$-$0056 & 19450 $\pm$ 270 & 7.00 $\pm$ 0.02 & 0.31 & \nodata & \nodata & \nodata & \nodata & \nodata \\
J2345$-$0102 & 33130 $\pm$ 450 & 7.20 $\pm$ 0.04 & 0.42 & \nodata & \nodata & \nodata & \nodata & \nodata \\
\enddata
\end{deluxetable}

\begin{deluxetable}{crccrcccc}
\tabletypesize{\footnotesize}
\tablecolumns{9}
\tablewidth{0pt} 
\tablecaption{Merger Systems in the ELM Survey}
\tablehead{
\colhead{Object}&
\colhead{$T_{\rm eff}$}&
\colhead{$\log g$}&
\colhead{$P$}&
\colhead{$K$}&
\colhead{Mass}&
\colhead{$M_2$}&
\colhead{$M_2(i=60^\circ)$}&
\colhead{$\tau_{\rm merge}$}\\
  & (K) &  & (days) & (km s$^{-1}$) & ($M_\odot$) & ($M_\odot$) & ($M_\odot$) & (Gyr)
}
\startdata 
J0022$-$1014 & 18980  & 7.15  & 0.07989 & 145.6 & 0.33 & $\geq 0.19$ & 0.23 & $\leq 0.73$ \\
J0755+4906   & 13160  & 5.84  & 0.06302 & 438.0 & 0.17 & $\geq 0.81$ & 1.12 & $\leq 0.22$ \\
J0818+3536   & 10620  & 5.69  & 0.18315 & 170.0 & 0.17 & $\geq 0.26$ & 0.33 & $\leq 8.89$ \\
J0822+2753   & 8880   & 6.44  & 0.24400 & 271.1 & 0.17 & $\geq 0.76$ & 1.05 & $\leq 8.42$ \\
J0849+0445   & 10290  & 6.23  & 0.07870 & 366.9 & 0.17 & $\geq 0.64$ & 0.88 & $\leq 0.47$ \\
J0923+3028   & 18350  & 6.63  & 0.04495 & 296.0 & 0.23 & $\geq 0.34$ & 0.44 & $\leq 0.13$ \\
J1053+5200   & 15180  & 6.55  & 0.04256 & 264.0 & 0.20 & $\geq 0.26$ & 0.33 & $\leq 0.16$ \\
J1233+1602   & 10920  & 5.12  & 0.15090 & 336.0 & 0.17 & $\geq 0.86$ & 1.20 & $\leq 2.14$ \\ 
J1234$-$0228 & 18000  & 6.64  & 0.09143 & 94.0  & 0.23 & $\geq 0.09$ & 0.11 & $\leq 2.69$ \\
J1436+5010   & 16550  & 6.69  & 0.04580 & 347.4 & 0.24 & $\geq 0.46$ & 0.60 & $\leq 0.10$ \\
J2119$-$0018 & 10360  & 5.36  & 0.08677 & 383.0 & 0.17 & $\geq 0.75$ & 1.04 & $\leq 0.54$ \\
NLTT 11748   & 8690   & 6.54  & 0.23503 & 273.4 & 0.18 & 0.76     & \nodata & 7.20 \\
\enddata
\tablecomments{This table is based on the data presented in \citet{kilic10b}, \citet{brown10a},
\citet{kilic10c}, and this paper. See \citet{mullally09}, \citet{steinfadt10}, and \citet{kawka10} for
additional observations of J1053+5200, J1436+5010, and NLTT 11748.}
\end{deluxetable}

\begin{figure}
\includegraphics[width=4.7in,angle=-90]{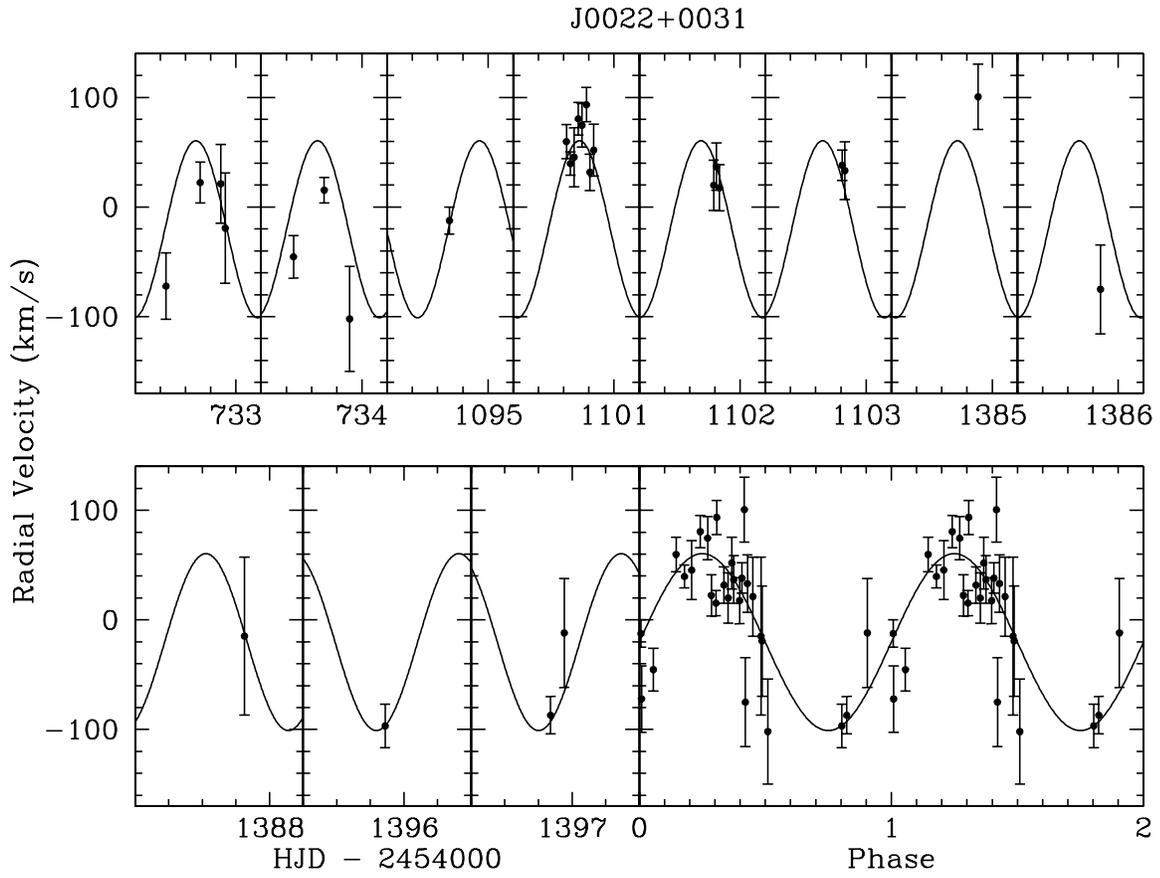}
\caption{Radial velocities and the best-fit orbit for J0022+0031.
The bottom right panel shows all of the data points phased
with the best-fit period.}
\end{figure}

\begin{figure}
\includegraphics[width=4.7in,angle=-90]{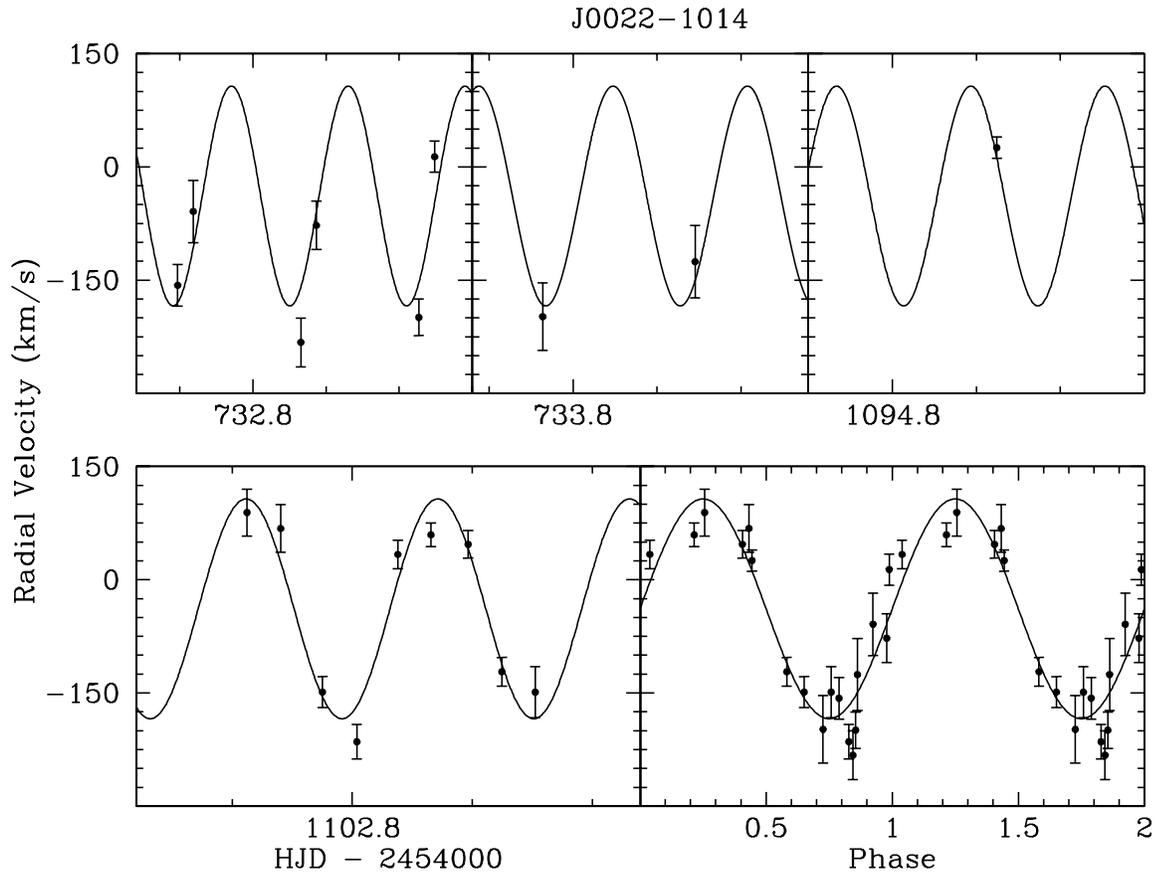}
\caption{Same as Figure 1, but for J0022$-$1014.}
\end{figure}

\begin{figure}
\includegraphics[width=4.7in,angle=-90]{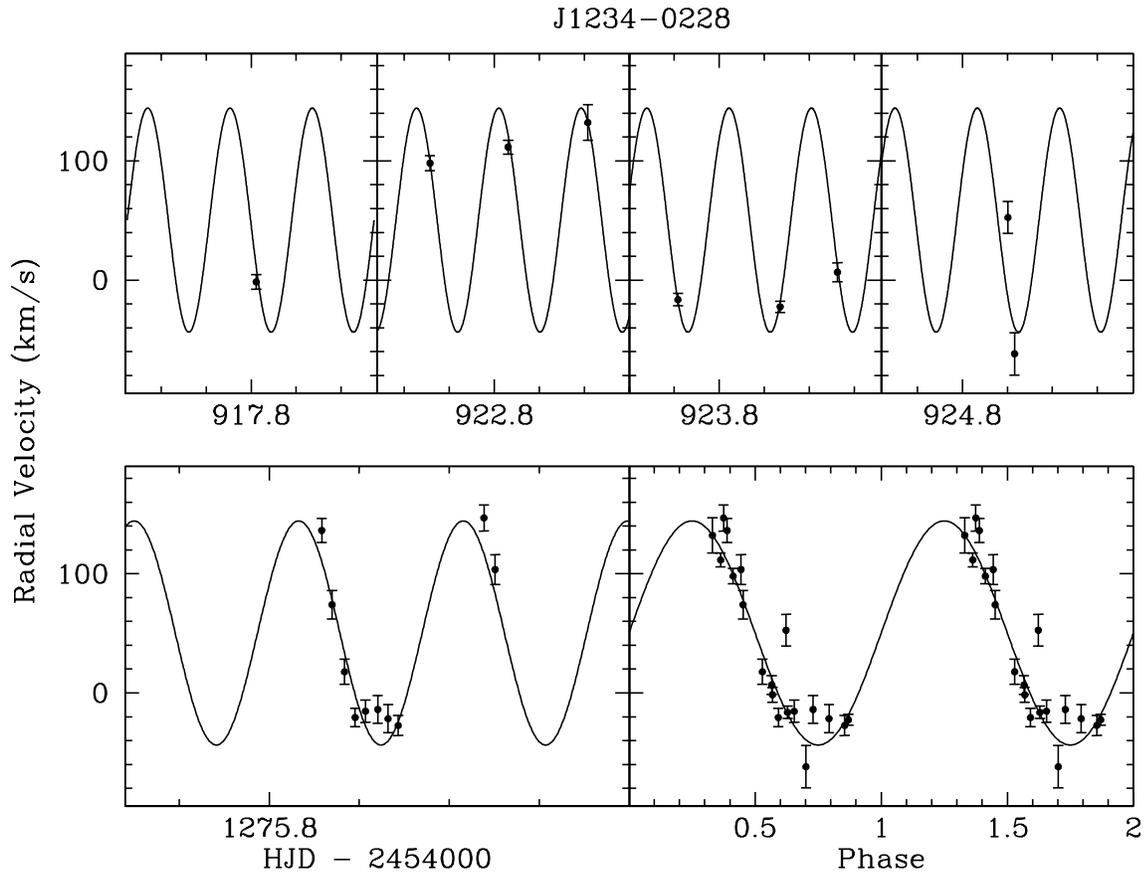}
\caption{Same as Figure 1, but for J1234$-$0228.}
\end{figure}

\begin{figure}
\includegraphics[width=4.7in,angle=-90]{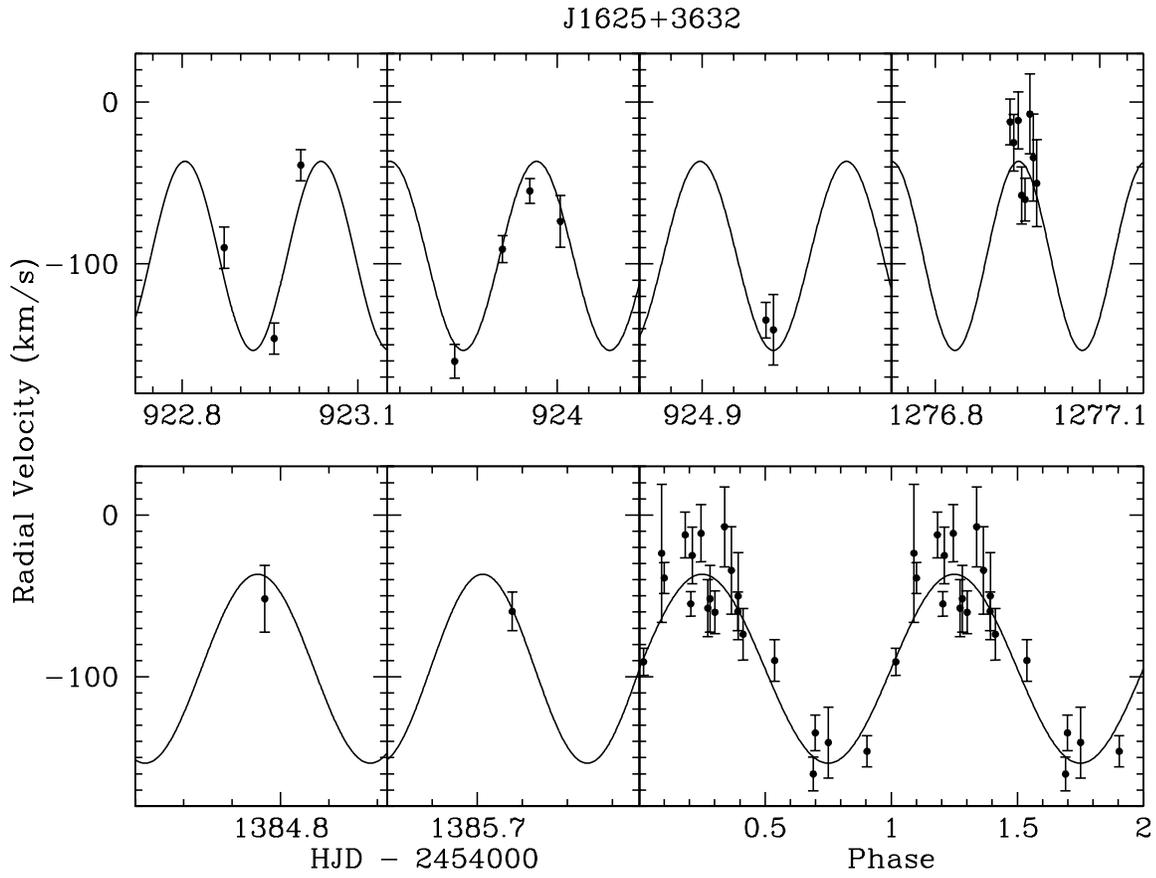}
\caption{Same as Figure 1, but for J1625+3632.}
\end{figure}

\begin{figure}
\includegraphics[width=4.7in,angle=-90]{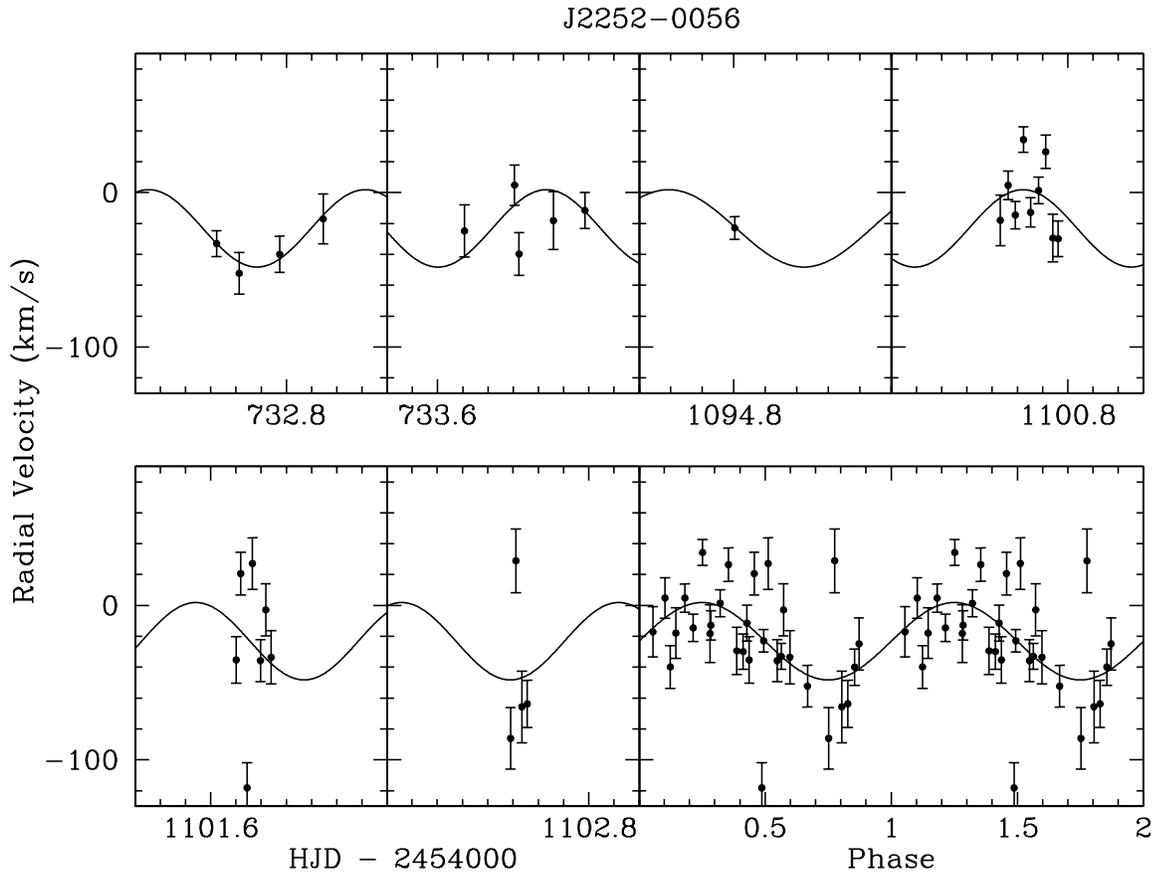}
\caption{Same as Figure 1, but for J2252$-$0056.}
\end{figure}

\begin{figure}
\includegraphics[width=4.7in,angle=-90]{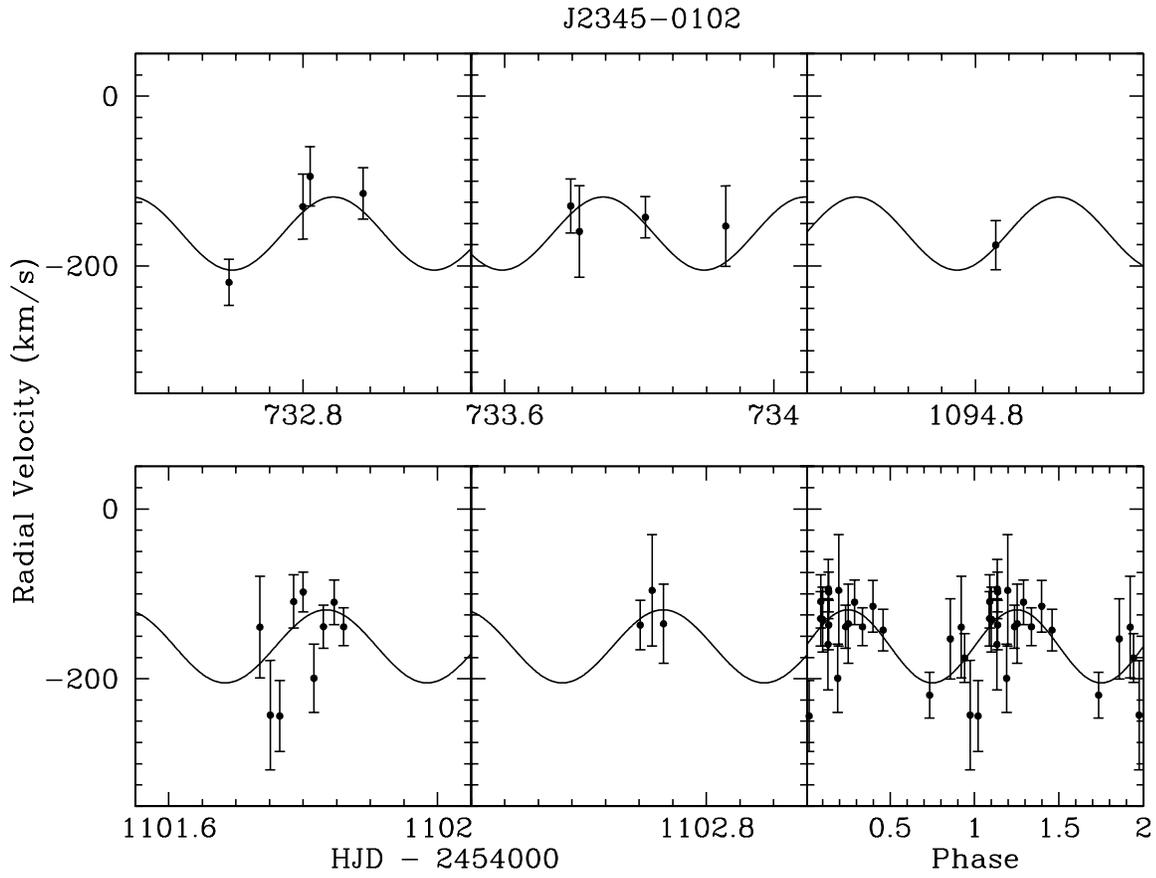}
\caption{Same as Figure 1, but for J2345$-$0102.}
\end{figure}

\begin{figure}
\plotone{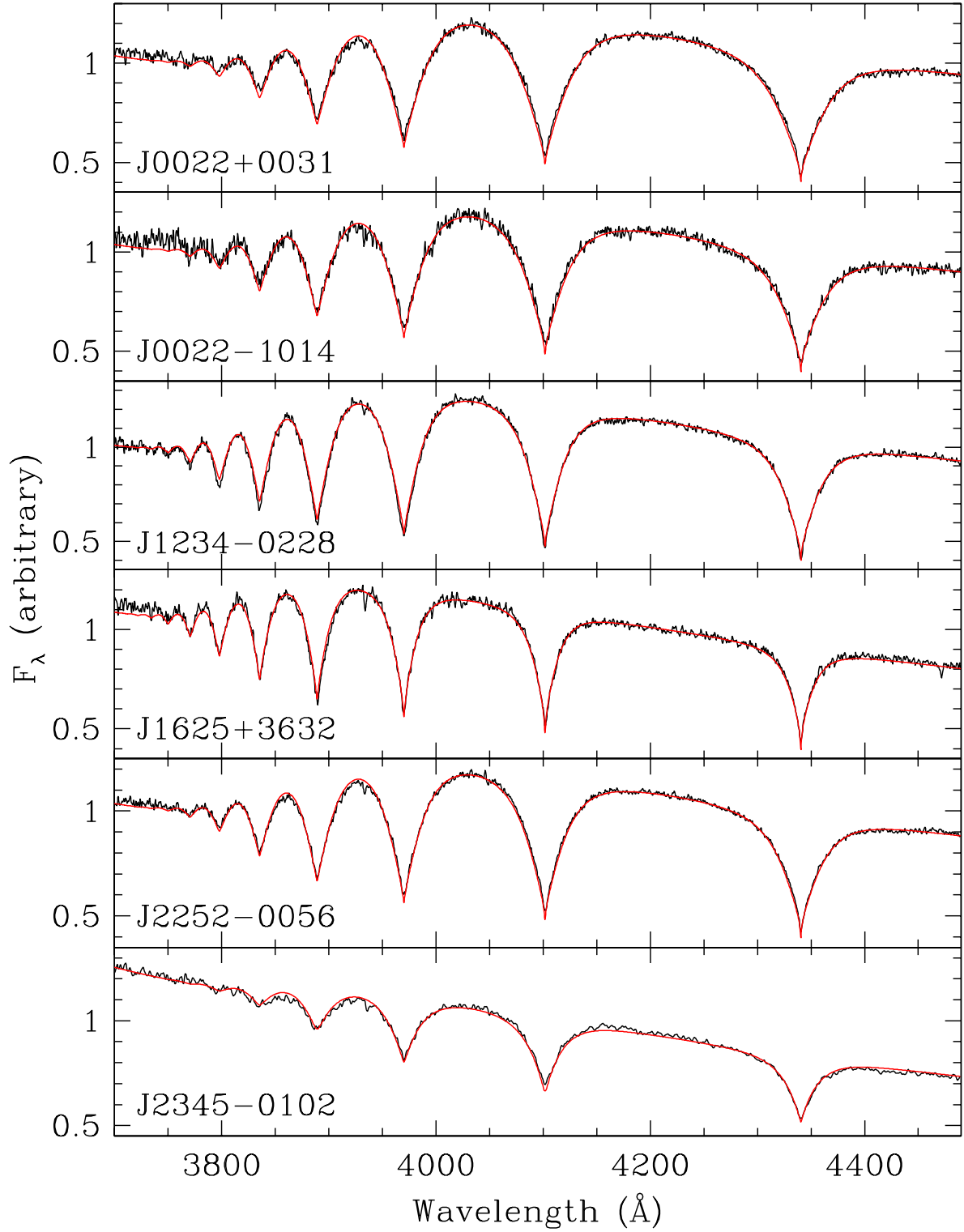}
\caption{Model fits (red solid lines) to the composite spectra of our targets (jagged lines).} 
\end{figure}

\begin{figure} 
\plotone{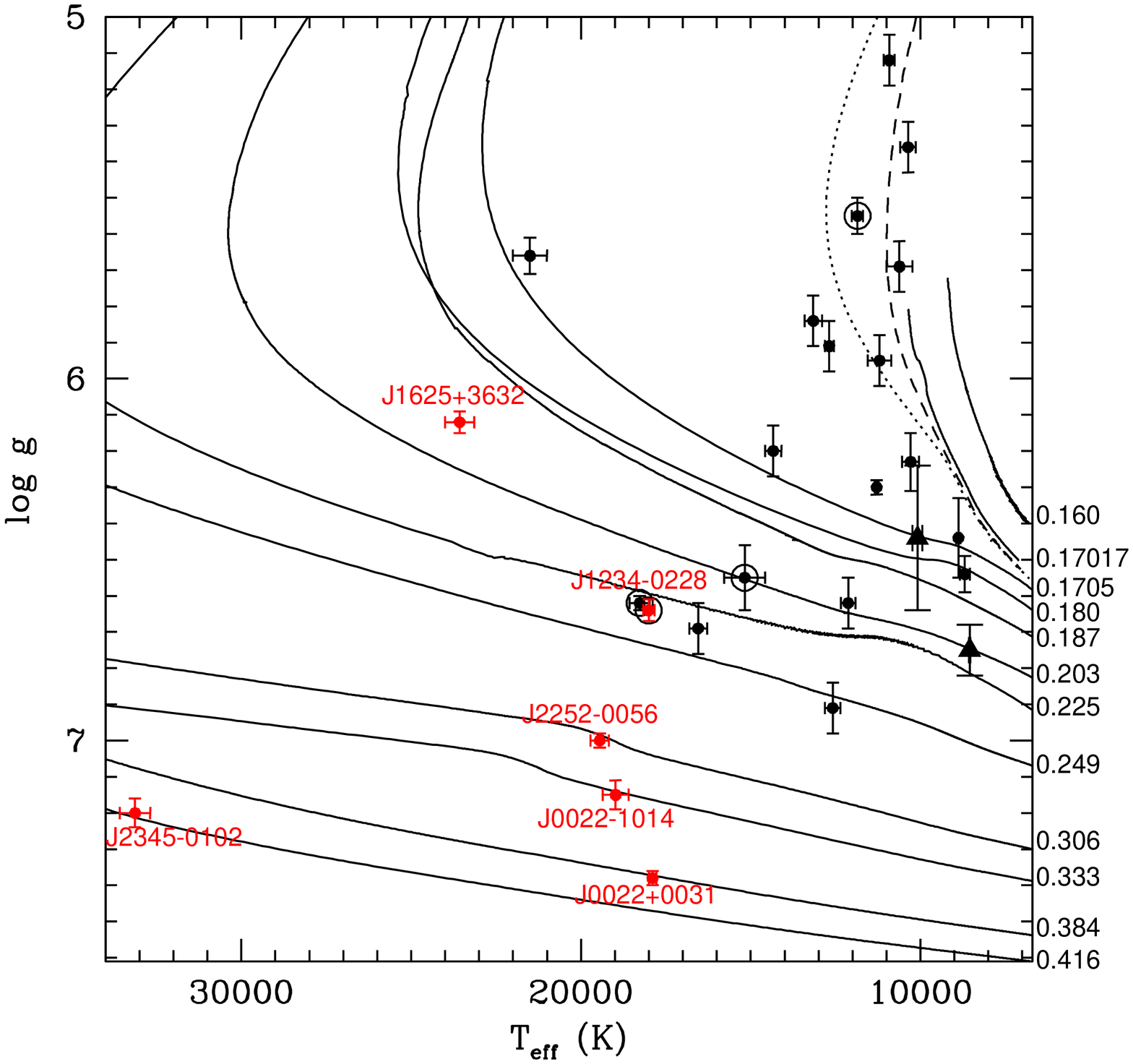}
\caption{The best fit solutions for the surface gravity and temperatures of our targets (red filled circles),
overlaid on tracks of constant mass from \citet{panei07} models updated by \citet{kilic10a}.
The dashed and dotted lines show solar metallicity and halo metallicity
(Z=0.001) models of \citet{serenelli01,serenelli02} for 0.17 $M_\odot$ WDs, respectively.
Spectroscopically confirmed WDs from the ELM survey and the sdB star HD 188112 \citep{heber03} are shown as
black circles. Companions to milli-second pulsars PSR J1012+5307 and J1911-5958A are shown as triangles.
The four objects with X-ray nondetections \citep[Table 1 and][]{agueros09b} are marked by open circles.}
\end{figure}

\begin{figure}
\includegraphics[width=4.7in,angle=-90]{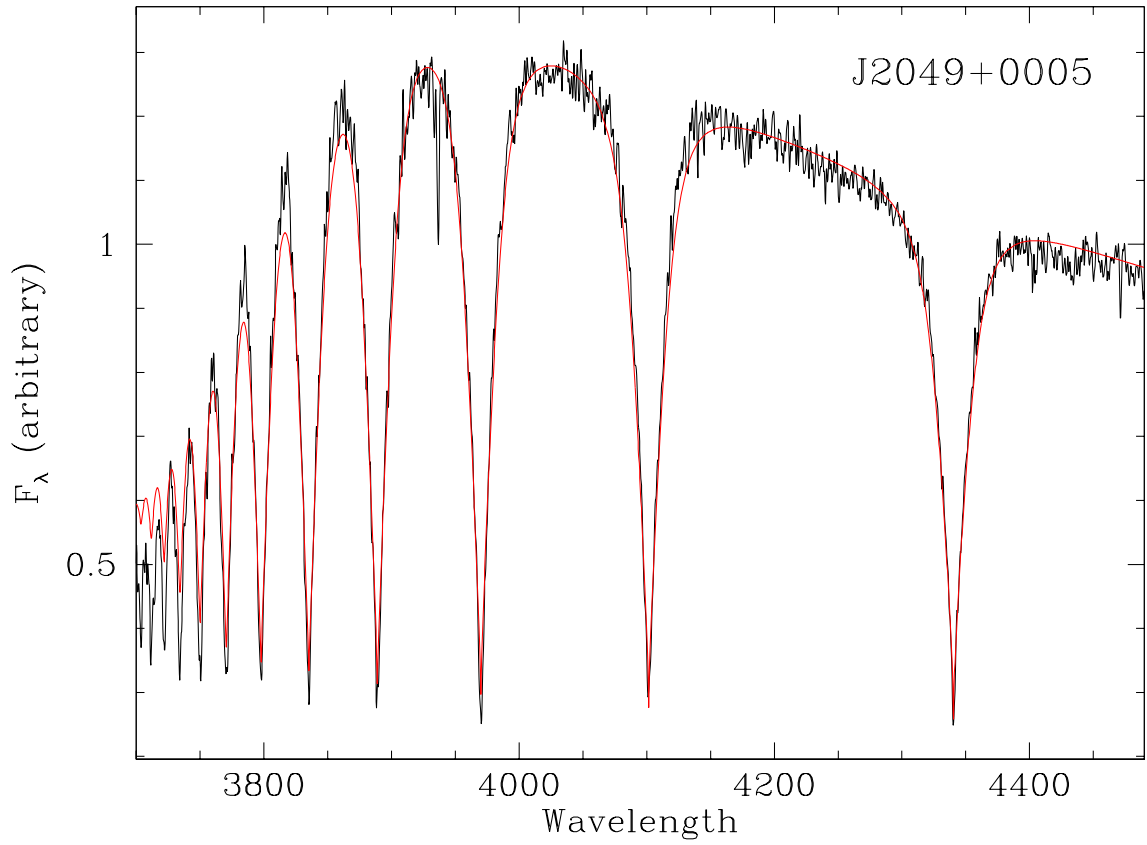}
\caption{Best-fit model (red solid line) to the observed spectrum of SDSS J2049+0005. The best-fit
model has $\log g=5$ and does not match the higher order Balmer lines. J2049+0005
is likely an A star.}
\end{figure} 

\begin{figure}
\plotone{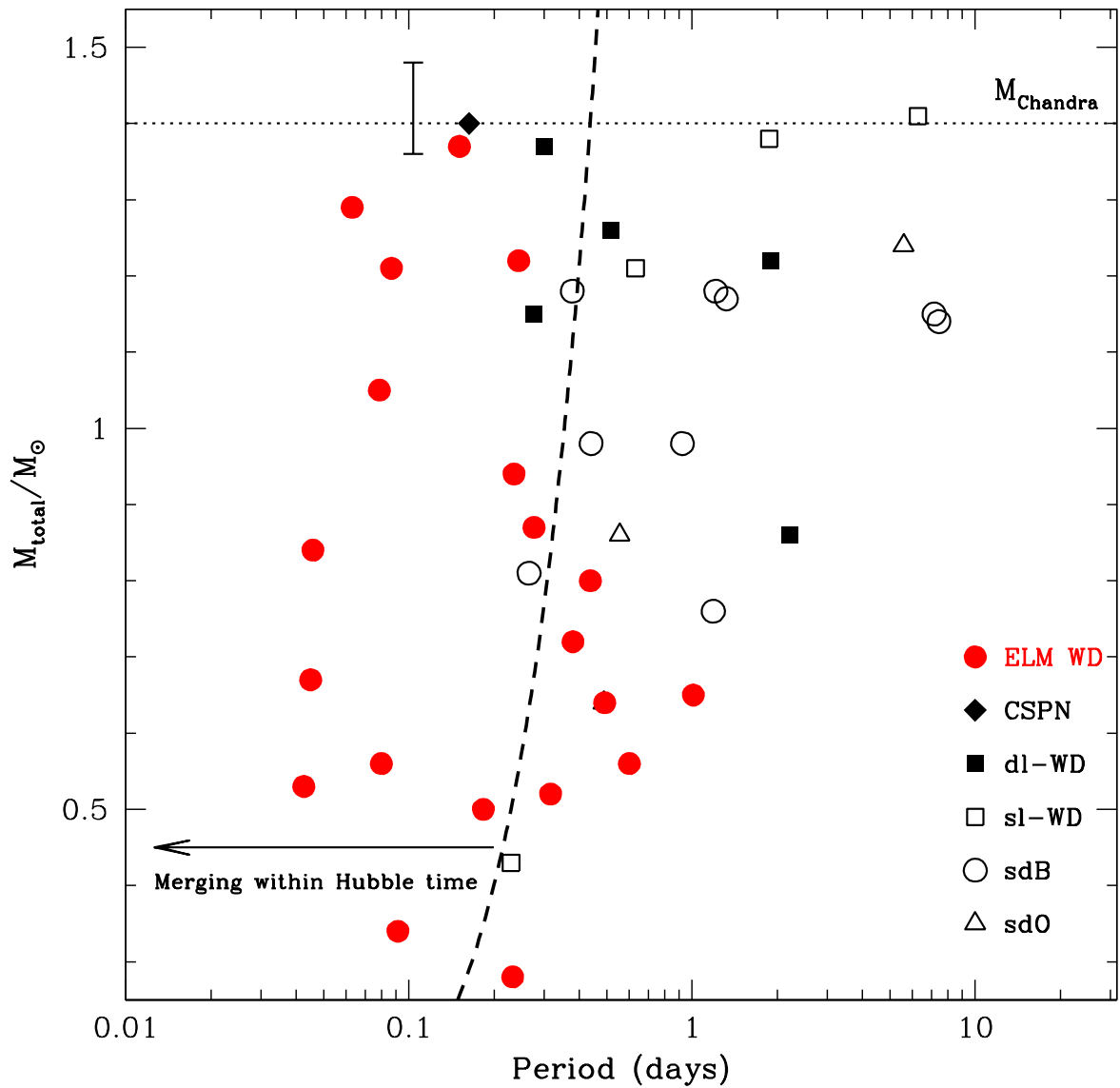}
\caption{Total mass plotted against logarithmic period of double degenerate systems
from the SPY survey \citep{geier10} and our ELM WD sample (red filled circles).
ELM WDs, central stars of planetary nebulae (CSPN), double-lined/single-lined WD systems,
and sdB/sdO systems are shown with different symbols.
\citet{geier10} assumes an inclination angle of 52$^{\circ}$ for the single-lined
systems, whereas we use an inclination angle of 60$^{\circ}$ for our systems.}
\end{figure}

\begin{figure}
\plotone{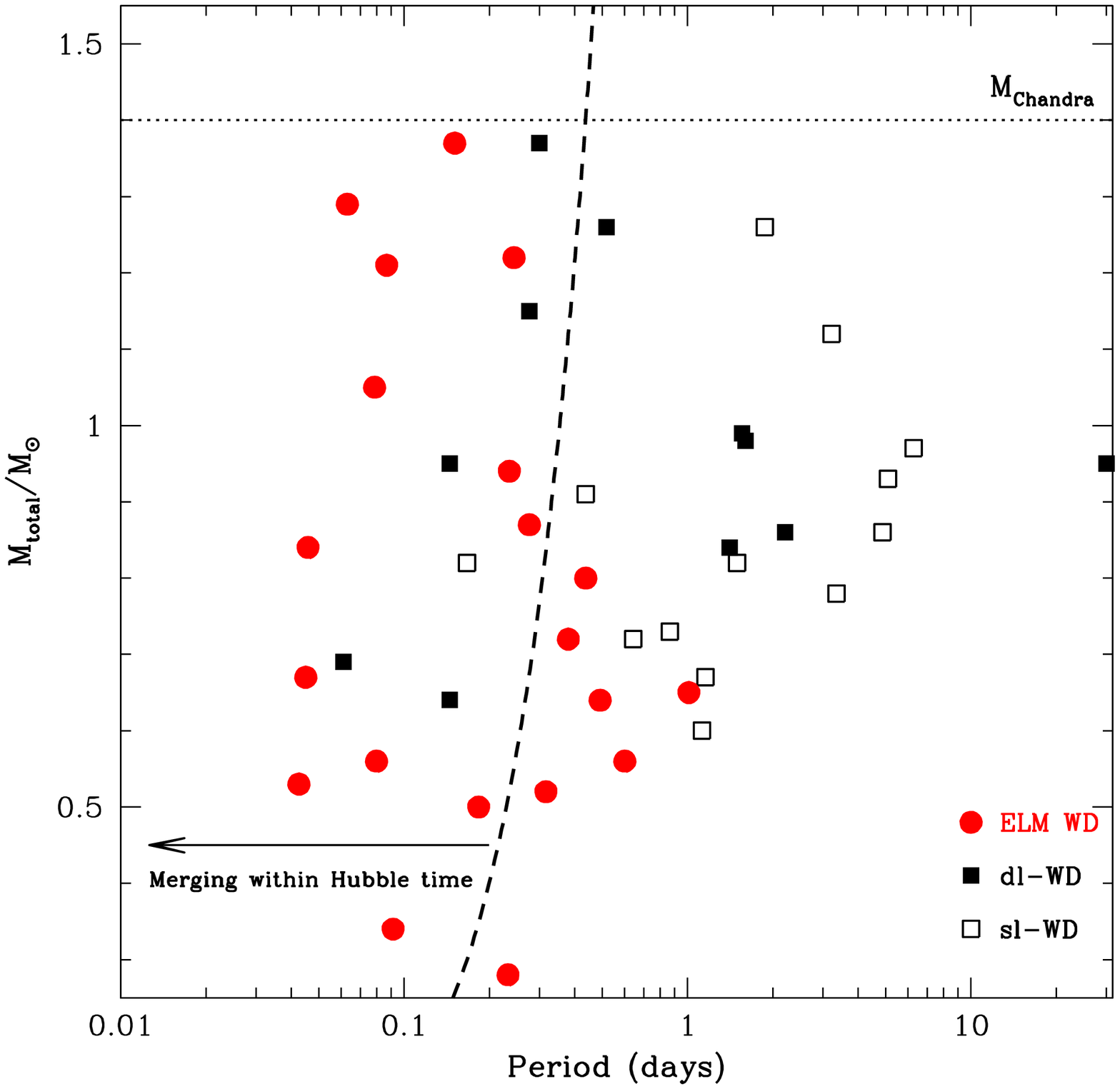}
\caption{Same as Figure 10, but including only the double WD systems found in the literature \citep{nelemans05,napiwotzki07}.}
\end{figure}

\begin{figure}
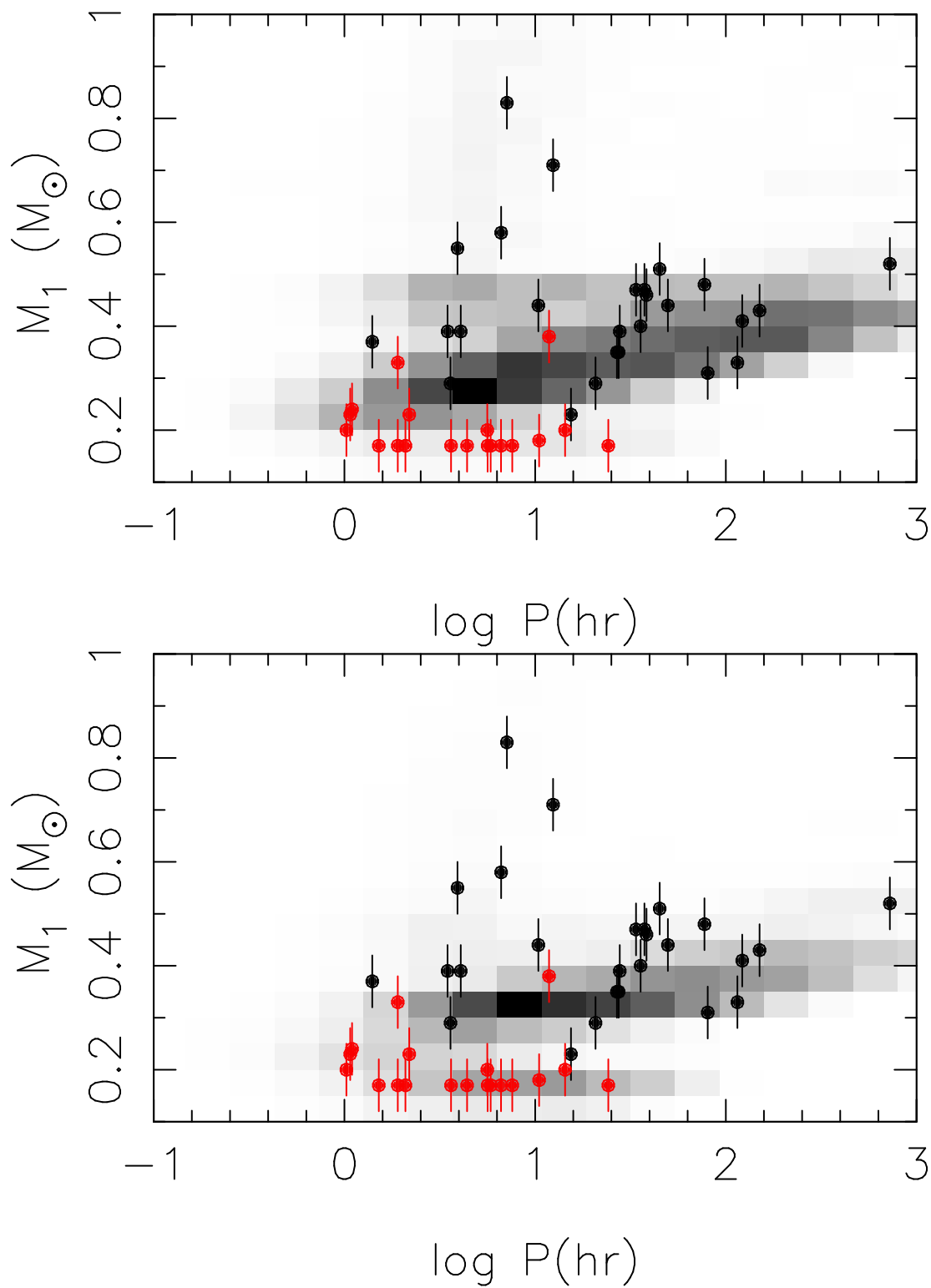

\includegraphics[width=4.0in,angle=-90]{f12a.ps}
\includegraphics[width=4.0in,angle=-90]{f12b.ps}
\caption{Model population of double WDs as a function of orbital period and mass
of the brighter WD of the pair (kindly made available by G. Nelemans). The top
panel uses \citet{hansen99} cooling models whereas the bottom panel uses modified \citet{driebe98}
cooling models. The observed binary WD population \citep{nelemans05} and our ELM WD sample (red filled circles)
are also shown.}
\end{figure}

\begin{figure} 
\plotone{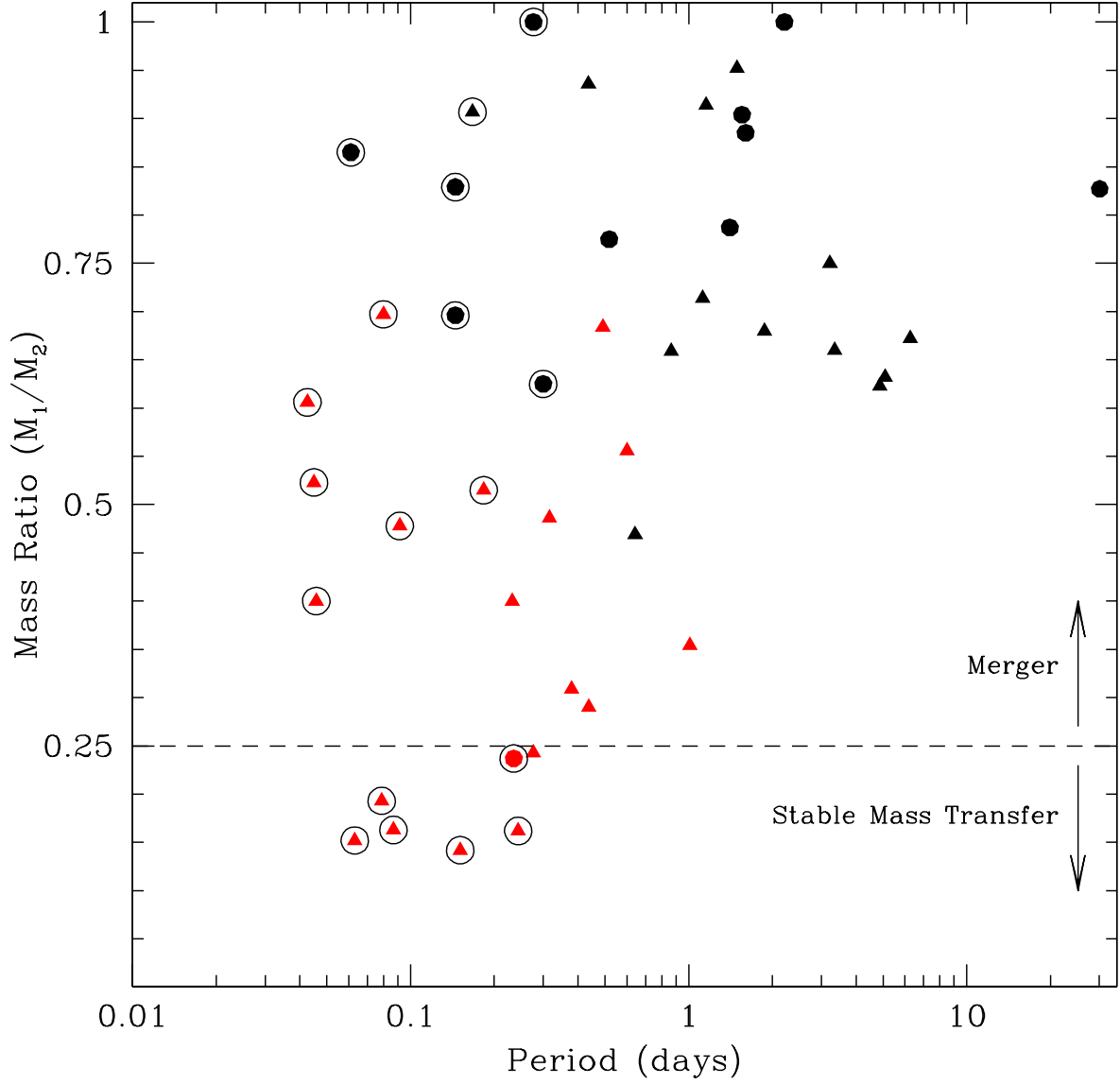} 
\caption{Mass ratios and periods for binary WD systems found in our ELM survey and the
literature. The mass ratio limit for stable mass transfer is also indicated.}
\end{figure} 

\clearpage
\appendix
\section*{APPENDIX A. Radial Velocity Data}

\begin{deluxetable}{crr}
\tabletypesize{\footnotesize}
\tablecolumns{3}
\tablewidth{0pt}
\tablecaption{Radial Velocity Measurements}
\tablehead{
\colhead{Object}&
\colhead{HJD}&
\colhead{$v_{helio}$}\\
 & $-$2450000 & (km s$^{-1}$)
}
\startdata
J0022+0031 &  732.722684 &  $-$72.2 $\pm$ 30.3 \\
\nodata    &  732.858519 &     22.4 $\pm$ 18.8 \\
\nodata    &  732.939955 &     21.2 $\pm$ 36.0 \\
\nodata    &  732.957466 &  $-$19.3 $\pm$ 50.4 \\
\nodata    &  733.728259 &  $-$45.4 $\pm$ 19.5 \\
\nodata    &  733.850239 &     15.3 $\pm$ 11.6 \\
\nodata    &  733.951397 & $-$102.0 $\pm$ 47.9 \\
\nodata    & 1094.846527 &  $-$12.4 $\pm$ 12.4 \\
\nodata    & 1100.810977 &     59.7 $\pm$ 15.6 \\
\nodata    & 1100.827123 &     39.7 $\pm$ 10.6 \\
\nodata    & 1100.841278 &     45.5 $\pm$ 26.9 \\
\nodata    & 1100.858385 &     80.6 $\pm$ 14.8 \\
\nodata    & 1100.872447 &     74.6 $\pm$ 19.8 \\
\nodata    & 1100.890352 &     93.5 $\pm$ 15.6 \\
\nodata    & 1100.904427 &     31.7 $\pm$ 16.6 \\
\nodata    & 1100.919566 &     51.9 $\pm$ 23.7 \\
\nodata    & 1101.895459 &     19.9 $\pm$ 22.9 \\
\nodata    & 1101.906050 &     36.9 $\pm$ 21.6 \\
\nodata    & 1101.917786 &     17.5 $\pm$ 21.0 \\
\nodata    & 1102.904928 &     38.2 $\pm$ 13.9 \\
\nodata    & 1102.915530 &     33.1 $\pm$ 26.3 \\
\nodata    & 1384.944307 &    100.6 $\pm$ 29.7 \\
\nodata    & 1385.929157 &  $-$75.1 $\pm$ 40.6 \\
\nodata    & 1387.925074 &  $-$14.8 $\pm$ 72.2 \\
\nodata    & 1395.943892 &  $-$96.8 $\pm$ 19.8 \\
\nodata    & 1396.936261 &  $-$87.0 $\pm$ 17.1 \\
\nodata    & 1396.976913 &  $-$11.9 $\pm$ 49.8 \\
J0022$-$1014 & 732.748459 & $-$156.9 $\pm$ 27.6 \\
\nodata & 732.759199  &  $-$59.1 $\pm$ 41.4 \\
\nodata & 732.832753  & $-$232.5 $\pm$ 32.4 \\
\nodata & 732.843424  &  $-$77.3 $\pm$ 32.2 \\
\nodata & 732.913470  & $-$199.3 $\pm$ 24.1 \\
\nodata & 732.924142  &     13.5 $\pm$ 20.7 \\
\nodata & 733.781989  & $-$198.3 $\pm$ 44.7 \\
\nodata & 733.872718  & $-$125.5 $\pm$ 47.9 \\
\nodata & 1094.862244 &     25.5 $\pm$ 14.1 \\
\nodata & 1102.756314 &     89.0 $\pm$ 31.1 \\
\nodata & 1102.770423 &     67.9 $\pm$ 31.6 \\
\nodata & 1102.787923 & $-$148.8 $\pm$ 20.6 \\
\nodata & 1102.801997 & $-$214.7 $\pm$ 22.9 \\
\nodata & 1102.818895 &     33.5 $\pm$ 18.9 \\
\nodata & 1102.832969 &     59.5 $\pm$ 15.6 \\
\nodata & 1102.848212 &     46.9 $\pm$ 18.2 \\
\nodata & 1102.862274 & $-$121.9 $\pm$ 19.0 \\
\nodata & 1102.876371 & $-$148.9 $\pm$ 33.8 \\
J1224$-$0228 & 917.805724 & $-$1.6 $\pm$ 6.2 \\
\nodata & 922.728604  &    98.1 $\pm$  6.4 \\
\nodata & 922.815478  &   111.6 $\pm$  5.8 \\
\nodata & 922.904031  &   132.2 $\pm$ 14.9 \\
\nodata & 923.754131  & $-$16.3 $\pm$  5.2 \\
\nodata & 923.867625  & $-$22.5 $\pm$  4.6 \\
\nodata & 923.931236  &     6.6 $\pm$  7.9 \\
\nodata & 924.850731  &    52.6 $\pm$ 13.4 \\
\nodata & 924.857965  & $-$61.9 $\pm$ 17.8 \\
\nodata & 1275.829077 &   136.3 $\pm$ 10.1 \\
\nodata & 1275.834841 &    73.9 $\pm$ 11.9 \\
\nodata & 1275.841936 &    17.8 $\pm$ 10.5 \\
\nodata & 1275.847700 & $-$20.6 $\pm$  7.8 \\
\nodata & 1275.853441 & $-$15.4 $\pm$  9.3 \\
\nodata & 1275.860270 & $-$13.8 $\pm$ 11.7 \\
\nodata & 1275.866022 & $-$21.6 $\pm$ 11.9 \\
\nodata & 1275.871763 & $-$27.3 $\pm$  8.4 \\
\nodata & 1275.919252 &   146.8 $\pm$ 11.0 \\
\nodata & 1275.925560 &   103.6 $\pm$ 12.5 \\
J1625+3632 &  922.872132 &  $-$90.0 $\pm$ 12.8 \\
\nodata    &  922.957182 & $-$146.1 $\pm$  9.7 \\
\nodata    &  923.002704 &  $-$39.0 $\pm$  9.6 \\
\nodata    &  923.837051 & $-$160.2 $\pm$ 10.4 \\
\nodata    &  923.912980 &  $-$90.8 $\pm$  8.4 \\
\nodata    &  923.956569 &  $-$54.9 $\pm$  7.6 \\
\nodata    &  924.004812 &  $-$73.7 $\pm$ 16.0 \\
\nodata    &  925.000912 & $-$134.7 $\pm$ 11.0 \\
\nodata    &  925.012835 & $-$140.7 $\pm$ 21.9 \\
\nodata    & 1276.936718 &  $-$12.2 $\pm$ 14.2 \\
\nodata    & 1276.943038 &  $-$25.0 $\pm$ 17.5 \\
\nodata    & 1276.951372 &  $-$11.3 $\pm$ 17.6 \\
\nodata    & 1276.957668 &  $-$57.6 $\pm$ 17.7 \\
\nodata    & 1276.964011 &  $-$60.2 $\pm$ 13.3 \\
\nodata    & 1276.972808 &   $-$7.3 $\pm$ 24.7 \\
\nodata    & 1276.979128 &  $-$34.2 $\pm$ 27.0 \\
\nodata    & 1276.985447 &  $-$50.1 $\pm$ 26.9 \\
\nodata    & 1384.783927 &  $-$51.8 $\pm$ 20.6 \\
\nodata    & 1385.738969 &  $-$59.6 $\pm$ 11.9 \\
\nodata    & 1396.823036 &  $-$23.7 $\pm$ 42.7 \\
J2252$-$0056 &  732.661721 &  $-$33.0 $\pm$  8.4 \\
\nodata      &  732.706500 &  $-$52.3 $\pm$ 13.5 \\
\nodata      &  732.786660 &  $-$40.0 $\pm$ 11.8 \\
\nodata      &  732.872943 &  $-$17.0 $\pm$ 16.2 \\
\nodata      &  733.653315 &  $-$24.9 $\pm$ 16.9 \\
\nodata      &  733.752698 &      4.8 $\pm$ 13.1 \\
\nodata      &  733.761692 &  $-$39.8 $\pm$ 13.9 \\
\nodata      &  733.829687 &  $-$18.2 $\pm$ 18.7 \\
\nodata      &  733.892232 &  $-$11.5 $\pm$ 11.7 \\
\nodata      & 1094.801506 &  $-$22.9 $\pm$  7.3 \\
\nodata      & 1100.666435 &  $-$18.0 $\pm$ 16.4 \\
\nodata      & 1100.681967 &      4.7 $\pm$  9.2 \\
\nodata      & 1100.696261 &  $-$14.5 $\pm$  8.9 \\
\nodata      & 1100.712232 &     34.4 $\pm$  8.3 \\
\nodata      & 1100.726399 &  $-$12.9 $\pm$  9.5 \\
\nodata      & 1100.742289 &      1.4 $\pm$  8.7 \\
\nodata      & 1100.756386 &     26.4 $\pm$ 10.8 \\
\nodata      & 1100.770645 &  $-$29.4 $\pm$ 15.4 \\
\nodata      & 1100.781270 &  $-$29.9 $\pm$ 11.5 \\
\nodata      & 1101.650617 &  $-$35.3 $\pm$ 15.1 \\
\nodata      & 1101.659471 &     20.6 $\pm$ 13.9 \\
\nodata      & 1101.672399 & $-$118.1 $\pm$ 16.3 \\
\nodata      & 1101.682989 &     27.1 $\pm$ 16.7 \\
\nodata      & 1101.698706 &  $-$35.8 $\pm$ 13.6 \\
\nodata      & 1101.709307 &   $-$2.8 $\pm$ 16.8 \\
\nodata      & 1101.719897 &  $-$33.6 $\pm$ 17.3 \\
\nodata      & 1102.645214 &  $-$86.1 $\pm$ 19.9 \\
\nodata      & 1102.655804 &     28.9 $\pm$ 20.7 \\
\nodata      & 1102.667655 &  $-$65.7 $\pm$ 23.2 \\
\nodata      & 1102.678245 &  $-$63.8 $\pm$ 15.3 \\
J2345$-$0102  &  732.689649 & $-$219.3 $\pm$ 27.1 \\
\nodata       &  732.799579 & $-$130.2 $\pm$ 38.3 \\
\nodata       &  732.810239 &  $-$94.4 $\pm$ 35.0 \\
\nodata       &  732.889312 & $-$114.6 $\pm$ 30.3 \\
\nodata       &  733.697952 & $-$129.3 $\pm$ 31.9 \\
\nodata       &  733.711030 & $-$159.5 $\pm$ 53.8 \\
\nodata       &  733.809108 & $-$142.7 $\pm$ 24.4 \\
\nodata       &  733.928533 & $-$153.0 $\pm$ 47.3 \\
\nodata       & 1094.830335 & $-$175.5 $\pm$ 28.8 \\
\nodata       & 1101.735791 & $-$139.1 $\pm$ 59.8 \\
\nodata       & 1101.751161 & $-$242.8 $\pm$ 64.4 \\
\nodata       & 1101.765224 & $-$243.8 $\pm$ 41.6 \\
\nodata       & 1101.786080 & $-$109.0 $\pm$ 31.3 \\
\nodata       & 1101.800154 &  $-$97.8 $\pm$ 23.5 \\
\nodata       & 1101.816149 & $-$199.4 $\pm$ 40.1 \\
\nodata       & 1101.830211 & $-$138.7 $\pm$ 25.3 \\
\nodata       & 1101.846137 & $-$109.9 $\pm$ 26.4 \\
\nodata       & 1101.860199 & $-$138.8 $\pm$ 22.3 \\
\nodata       & 1102.701680 & $-$136.8 $\pm$ 29.2 \\
\nodata       & 1102.719307 &  $-$95.9 $\pm$ 65.7 \\
\nodata       & 1102.736402 & $-$135.1 $\pm$ 46.7 \\
\enddata
\end{deluxetable}

\end{document}